\documentclass[manuscript,anonymous,acmsmall]{acmart}
\usepackage{xcolor}
\usepackage{graphicx}
\usepackage{float}
\usepackage{tcolorbox}
\usepackage{rotating}
\usepackage{multirow}
\usepackage{longtable}
\usepackage{pdflscape}
\usepackage[utf8]{inputenc}
\usepackage{adjustbox}
\usepackage{array}
\usepackage{pdflscape}
\usepackage{subcaption}

\AtBeginDocument{%
  \providecommand\BibTeX{{%
    \normalfont B\kern-0.5em{\scshape i\kern-0.25em b}\kern-0.8em\TeX}}}

\newif\ifcomments

\newcommand{\shirin}[1]{\ifcomments{\bf \textcolor{purple}{Shirin: #1}}\fi}

\newcommand{\sayak}[1]{\ifcomments{\bf \textcolor{blue}{Sayak: #1}}\fi}

\acmConference[CSCW '24]{CSCW '24: ACM Conference on Computer-Supported Cooperative Work and Social Computing}{November 9--13, 2024}{San Jose, Costa Rica}

\begin{document}

\title[Gender-Based Hate Speech \#MeToo]{Exploring Gender-Based Toxic Speech on Twitter in Context of the \#MeToo movement: A Mixed Methods Approach}

\author{Sayak Saha Roy}
\affiliation{%
  \institution{The University of Texas at Arlington}
  \country{USA}
}
\email{sayak.saharoy@mavs.uta.edu}

\author{Ohad Gilbar}
\affiliation{%
  \institution{The Hebrew University of Jerusalem}
  \country{Israel}
}
\email{ohad.gilbar@mail.huji.ac.il}

\author{Christina Palantza}
\affiliation{%
  \institution{Vrije Universiteit Amsterdam}
  \country{Netherlands}
}
\email{c.palantza@vu.nl}

\author{Maxine Davis}
\affiliation{%
  \institution{Rutgers University}
  \country{USA}
}
\email{maxine.davis@rutgers.edu}

\author{Shirin Nilizadeh}
\affiliation{%
  \institution{The University of Texas at Arlington}
  \country{USA}
}
\email{shirin.nilizadeh@uta.edu}

\renewcommand{\shortauthors}{Anonymous Author(s)}

\begin{abstract}
The \#MeToo movement has catalyzed widespread public discourse surrounding sexual harassment and assault, empowering survivors to share their stories and holding perpetrators accountable. While the movement has had a substantial and largely positive influence, this study aims to examine the potential negative consequences in the form of increased hostility against women and men on the social media platform Twitter. By analyzing tweets shared between October 2017 and January 2020 by more than 47.1k individuals who had either disclosed their own sexual abuse experiences on Twitter or engaged in discussions about the movement, we identify the overall increase in gender-based hostility towards both women and men since the start of the movement. We also monitor 16 pivotal real-life events that shaped the \#MeToo movement to identify how these events may have amplified negative discussions targeting the opposite gender on Twitter. Furthermore, we conduct a thematic content analysis of a subset of gender-based hostile tweets, which helps us identify recurring themes and underlying motivations driving the expressions of anger and resentment from both men and women concerning the \#MeToo movement. This study highlights the need for a nuanced understanding of the impact of social movements on online discourse and underscores the importance of addressing gender-based hostility in the digital sphere. 
\end{abstract}




%

\maketitle
\section{Introduction}
\subsection{Origins and Impact of the \#MeToo Movement}
The \#MeToo movement was introduced in 2006 by social activist Tarana Burke, but the movement only gained traction more than 11 years later in October 2017, when the Harvey Weinstein scandal~\citep{weinsteinscandal} prompted actress Alyssa Milano to encourage women to post their sexual harassment/ abuse experiences on the micro-blogging website Twitter. The \#MeToo movement has been a monumental force in raising awareness~\citep{masciantonio2020sexual} changing norms on sexual harassment and assault, empowering survivors to share their stories and demanding justice~\citep{mendes2018metoo}. The online discourse of the \#MeToo movement has profoundly impacted public opinion, workplace policies, and legal outcomes~\citep{fileborn2019}. The online discourse surrounding the \#MeToo movement highlights the power of digital platforms in shaping societal attitudes and driving tangible progress in addressing sexual harassment and assault. 

However, the movement is also largely criticized. There are widely endorsed negative attitudes towards it with several surveys~\citep{roden2022metoo,kunst2019sexism} showing that individuals, especially men of certain ideologies, view the movement negatively and express hostility and concerns about false accusations and the potential for unjust damage to their reputation~\citep{maricourt2022metoo}, leading to the creation of the \#NotAllMen counter-narrative, which seeks to differentiate men who are not involved in sexual misconduct~\citep{pettyjohn2019howiwillchange,segraves2022notallmen}. As a result, some men may adopt increasingly misogynistic attitudes and behaviors as a form of resistance or retaliation, further perpetuating gender inequality and undermining the progress made by the \#MeToo movement~\citep{brake2019coworker,flood2019men}. Simultaneously, the increased visibility of sexual misconduct cases and the sheer number of women sharing their experiences have, in some cases, led to a heightened sense of mistrust and anger towards men~\citep{maricourt2022metoo} by women who may exacerbate existing gender tensions and contribute to a hostile atmosphere~\citep{bhattacharya2014inferring}. 

\subsection{Toxic Speech and the \#MeToo Movement}

Most of the public discourse on \#MeToo has taken place online, where opinions are usually expressed fragmented and polarized~\citep{suovilla2020metoo}, and indeed online toxic speech has been identified as a problem of \#MeToo by theoretical approaches~\citep{lindgren2019movement}, and this has been confirmed by a qualitative analysis of \#MeToo related Tweets~\citep{goel2020understanding}. Toxic speech appears to escalate to a considerable level, as the anonymity of online platforms allows for the spread of doxing and other forms of online harassment, possibly leading to real-life consequences such as job loss and physical harm~\citep{citron2014hate}. In addition, false reporting rates for sexual assault are comparable to those for other types of crimes~\citep{PewResearch2017}, while women who have spoken out against sexual violence have faced harassment, threats, and retaliation in real life~\citep{mendes2019digital}. 

\subsection{Existing Qualitative and Quantitative Analyses of the \#MeToo Movement}

As pointed out by Suovilla et al.~\citep{suovilla2020metoo}, it is crucial to comprehend the impact of such digital media-based movements on society. To work towards this, we interpret the \#MeToo movement as follows; the bold opposing views towards \#MeToo that are expressed as toxic speech underline the turbulence it has brought to society. This could change the social processes that are part of the current social concept of gender, as in Risman et al.' s~\citep{risman2018gender} theoretical model of gender as a social structure. This turbulence could signify the movement pointing to an array of cognitive biases about the two genders that the current ideology and organizational practices have fostered (such as men being in charge, women being submissive to men, etc.). Since these cognitive biases have not yet been overcome, the extreme toxic reactions could be an early manifestation of a change process, especially when expressed by those suffering and victimized by the current status quo. This interpretation should be further explored by collecting and analyzing data from the actual discourse around the \#MeToo movement where it began and primarily evolves, i.e., on the Twitter social media platform. This enables the formulation of a more accurate theoretical interpretation of the movement through an approach closer to the widely used form of qualitative analysis of grounded theory~\citep{strauss1997grounded,charmaz2010sage}, which is an inductive, iterative approach to construct theory by interacting with the data and thus performing an empirical analysis~\citep{charmaz2008grounded}. 

There have been quite a few qualitative analyses of \#MeToo posts. Drewett et al.\cite{drewett2021breaking} analyzed Tweets from the first day of the movement, which were mostly disclosures and revealed differences in power between perpetrators and victims. Others have studied the movement's spread, origins, its impact on policy and legislation, e.g., Loney-Howes et al.~\citep{loney2020online}, as well as its inter-sectional aspects \cite{trott2021networked}. Goel et al.~\cite{goel2020understanding} specified that the exact topics of the\#MeToo Tweets are related to gender equality, feminism, the spread of the movement, and doubting the accusations. An interesting finding of Goel et al. ~\cite{goel2020understanding} on toxic speech is its link with racial inequalities. More specifically, people of color see the movement as an avenue to address them, while white individuals approach the movement more politically ~\cite{mueller2021demographic}. Bogen et al.~\cite{bogen2019metoo,bogen2021metoo, bogen2022sexual} found that social responses to disclosures are generally positive and supportive, but men appear more reluctant to disclose sexual harassment/abuse. This was corroborated by Lowenstein~\citep{lowenstein2021me}, and the views of the men disclosing were both of hegemonic and non-hegemonic masculinity. The study closest to the objectives of the present one was that of Nutbeam et al.~\cite{nutbeam2022negative}, which analyzed Tweets collected specifically from a period with several \#MeToo events (June 2019) and found that negative views of the movement included doubting accusations and the impact on the accused.

Despite the prominence of toxic speech linked with the movement and its implications, there is no qualitative work focusing specifically on the themes in offensive posts targeting the opposite gender. Examining that can provide valuable insights into the factors that drive toxic discourse and how it can be addressed more effectively and point towards healthier ways to promote societal change. Quantitative research on the association of the movement with toxic speech is also lacking. Furthermore, the movement was ignited by a real-life event, i.e. the accusations of sexual misconduct by Harvey Weinstein, and the hype was maintained by subsequent real-life events. To the best of our knowledge, there have been no quantitative studies regarding the association between the movement and an increase in hatred towards the opposite gender or the extent to which real-life \#MeToo events are associated with increased toxic posts towards the opposite gender on social media platforms. Studying these connections can reveal patterns and triggers for toxic speech.

\subsection{The present study} 

This work addresses these gaps by exploring the relation of the \#MeToo movement to the prevalence, nature, and themes of toxic speech between men and women on Twitter. 
To achieve this, we formulate the following three research questions:


\begin{tcolorbox}[ 
    colframe=black,  
    boxrule=0.7pt,
    sharp corners,
    left=2mm, 
    right=2mm, 
    top=2mm, 
    bottom=2mm 
]

\textbf{RQ1: }Do victims who disclose their sexual harassment experiences exhibit increased tendencies to share toxic posts against the opposite gender following their experience?
\end{tcolorbox}
We first investigate if the online community's response to such disclosures differs based on the gender of the victim, identifying any disparities in support from the online community.
We then explore if victims demonstrate an increased propensity to share toxic content against the opposite gender, which may indicate the extent of psychological and emotional consequences of their traumatic experiences \textbf{(Section~\ref{disclosure-analysis})}.

\begin{tcolorbox}[ 
    colframe=black,  
    boxrule=0.7pt,
    sharp corners,
    left=2mm, 
    right=2mm, 
    top=2mm, 
    bottom=2mm 
]
\textbf{RQ2: }Is there an association between the \#MeToo movement and an increase in toxic tweets targeting the opposite gender?
\end{tcolorbox}

We evaluate tweets shared by a random sample of 10k female and 10k male users to determine if they were more likely to share toxic posts against the opposite gender after the start of the \#MeToo movement \textbf{(Section~\ref{toxic-opposite-gender})}.

\begin{tcolorbox}[
    colframe=black,  
    boxrule=0.7pt,
    sharp corners,
    left=2mm, 
    right=2mm, 
    top=2mm, 
    bottom=2mm 
]
\textbf{RQ3: }To what extent were real-life \#MeToo events associated with an increase in toxic posts towards the opposite gender?
\end{tcolorbox}

Our study identifies 16 significant real-life events that occurred between October 2017 and January 2020 that contributed majorly to the development of the \#MeToo movement, and we determine how they influenced the prevalence of gender-based toxic posts on Twitter \textbf{(Section ~\ref{real-life_events_analysis})}.

\begin{tcolorbox}[
    colframe=black,  
    boxrule=0.7pt,
    sharp corners,
    left=2mm, 
    right=2mm, 
    top=2mm, 
    bottom=2mm 
]
\textbf{RQ4: }What are the dominant themes in toxic posts targeting the opposite gender that were influenced by the \#MeToo movement and how do these themes interact with one another?
\end{tcolorbox}
In the context of the \#MeToo movement, we perform a qualitative evaluation of a selection of toxic tweets targeted towards the opposite gender, shared by both female and male users. Our analysis aims to delineate not only the explicit themes that emerge within these posts but also to investigate how these themes interact and coexist.
This allows us to have a deeper comprehension of the evolving landscape of societal dialogue, illuminating the direction and nature of these changes \textbf{(Section~\ref{thematic_analysis})}.

\section{Data collection and processing}
\label{data_collection}
In order to investigate our research questions, we designed a methodology encompassing data collection and the employment of various quantitative and qualitative analysis models. 
Figure~\ref{framework_pic} provides a visual outline of our study's framework and procedural approach. \shirin{briefly explain the visual unless you explain it in other sections by explicitly referring to it.}\sayak{We refer to what we are doing for each RQ in the paragraphs below, which can be looked up on the image, so I think its fine?} \shirin{it's not exactly the same but let's submit. remember to add later.}
Our research framework entails several integral steps, which are described as follows:

\begin{figure*}
\centering
\includegraphics[width=\textwidth]{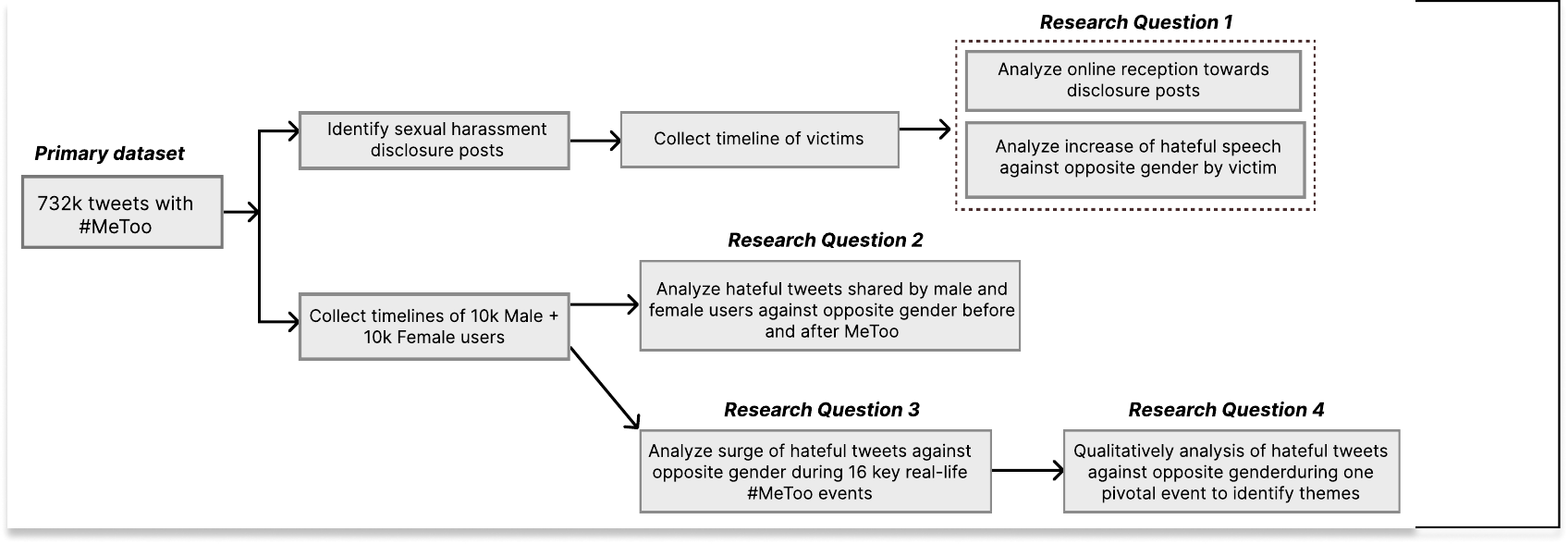}
\caption{Workflow of our analysis for RQ1 through RQ4}
\label{framework_pic}
\end{figure*}

\subsection{Primary dataset: Collecting \#MeToo tweets} 
Utilizing the Twitter API version 1.1~\citep{twitter-api11}, we gathered a pool of 732K tweets featuring the keyword hashtag `\#MeToo' from October 2017 to January 2020. Given the nature of the API, we were able to obtain a broad and randomized 1\% sample. We then filtered out retweets and replies to focus our analysis on original tweets connected to the `\#MeToo' movement. This process enabled us to concentrate on the fundamental content being propagated and crafted within this sociocultural discourse. We also focused on English tweets for our analysis, filtering them using both Twitter's inbuilt `en' language filter, followed by the Python library, langdetect~\citep{langdetect}, leaving us with a sample of 588K tweets. Intriguingly, these tweets were composed by nearly as many unique users (584.7K), a testament to the extensive reach and personal engagement spurred by the `\#MeToo' phenomenon. This dataset, serving as our primary corpus, laid the groundwork for various subsequent analyses addressing our research queries.


\subsection{Identification of the User’s Perceived Gender}
A major part of our analysis is heavily focused on determining and comparing the toxic characteristics of tweets shared by male and female users who participated in the \#MeToo movement.
Since it is not mandatory for users to disclose their gender on Twitter, we utilized the methodology introduced by Nilizadeh et al.~\citep{nilizadeh2016twitter} for identifying \textit{perceived gender} using the Face++ API~\citep{faceplusplus:2020} and 92,626 unique first names and their gender profiles in the U.S. from 1990 to 2013~\citep{mislove2011understanding}. 
Also, users have increasingly adopted the use of gender-identifiable pronouns on their social media profiles including Twitter~\cite{wareham2020pronouns,bottomline}, which provided us with a great opportunity to identify their gender as self-disclosed by the user. We thus searched for these gender-identifiable terms, such as ‘he/his,’ ‘she/her,’ ‘they,’ etc., in the profile description of the users, since it indicates the gender that the user wants to be identified as online~\citep{wareham2020should}. 
This data when available was given the highest preference among the three approaches.
To check the effectiveness of our method, three independent coders manually evaluated a sample of 200 random accounts from our dataset and identified the perceived gender by looking at their usernames, full names, profile pictures, and profile descriptions. 
The agreement between the algorithm and the coders after majority voting was 95.5\%, with a Cohen's kappa of 0.83 between the coders which shows substantial agreement between the coders. 
Using this algorithm, we were able to identify 353,815 users as either male or female, which was around 60\% of our dataset. This volume is comparable 
with that of Nilizadeh et al's~\cite{nilizadeh2016twitter} work (Who were able to identify the gender of about 57~\% of users in their dataset). 
\shirin{you should say if this is good enough and how it compares with the previous works/ approaches? I remember in our paper, we could identify gender for about 50-60 percent of users, so you can refer to that finding and say this approach could identify the same amount or more, etc.}~\sayak{Added}
To respect users' privacy settings and only use tweets that are publicly available, we eliminated any protected accounts and consider a total of 352,978 (215,408 female and 147,570 male) users for our analysis, who authored 354,249 tweets in our dataset.
We decided to omit gender-ambiguous users (which comprised nearly 40\% of all users in our dataset) so as to maintain high precision in our gender categorization. This precision is crucial for our analysis since it depends on contrasting the viewpoints between male and female users towards their respective opposite gender during the \#MeToo movement. However, it is important to acknowledge this approach as a potential limitation of our work, as it may exclude non-binary or transgender individuals from our analysis. Also, despite our best efforts, it is possible that some non-binary individuals may have been incorrectly categorized, thus adding a small amount of noise to our dataset.
We also filtered out potential bot accounts from our dataset using the Botometer API~\citep{botometer}, which checks various attributes of a Twitter account, such as posting patterns, their user community network, identifiable information shared, etc. and generates a score that indicates whether the account is a bot or a real user. 
Based on the observation of the authors of this tool, they finalized 0.43 as a suitable score for filtering out bot accounts. Using this threshold, we eliminated 39.4K accounts suspected of being bots.

\subsection{Identification of Self-disclosure Posts}
In \textbf{RQ1}, we aim to study tweets where individuals disclose their sexual harassment experiences. 
To collect such posts from Twitter, we first used a set of 21 seed keywords that are prominent in self-disclosure text, as identified by Roden et al.~\citep{roden2022metoo}. 
Three independent coders manually went through a random sample of 1K tweets that were using at least one of these keywords, and in the process, identified 7 more phrases that indicated self-disclosure: ‘he grabbed my,’ ‘she grabbed my,’ ‘abused me,’ ‘tied me,’ ‘he touched himself,’ ‘he touched my,’ and ‘she touched my.’ 
Using these keywords coupled with the term '\#MeToo', we found 27,930 tweets as self-disclosures, which is around 7.9\% of the tweets in our dataset.
17,567 (about 63\%) of them were posted by 16,984 unique female users compared to 10,363 (about 37\%) posted by 10,127 male users. 
To identify the effectiveness of this method, our coders evaluated a random sample of 200 tweets marked as `self-disclosure' and found a false positive rate of only 8\%, outperforming a previous implementation (14\%)~\citep{gallagher2019reclaiming}. However, this margin of error does not undermine the validity of our findings. Our examination revealed that close to 63\% of the posts originated from females, with males accounting for 37\%. Even after allowing for an 8\% uncertainty, the gender gap in our data remains prominent.
\shirin{what was the fp of the previous implementation? Explain how this does not affect our findings.}\sayak{Added}

\begin{table*}[!htb]
\centering
\caption{Descriptive statistics for users who posted disclosures about their harassment experience. }
\label{descriptive-disclosure}
\resizebox{\textwidth}{!}{%

\begin{tabular}{lccccc|ccccc}
\hline \\[-1.8ex] 
& \multicolumn{5}{c}{Male users (n=10,363)} & \multicolumn{5}{c}{Female users (n=16,984)} \\
\hline \hline \\[-1.8ex] 
Statistic & \multicolumn{1}{c}{Type} & \multicolumn{1}{c}{Min} & \multicolumn{1}{c}{Max} & \multicolumn{1}{c}{Mean}&\multicolumn{1}{c}{Median} & \multicolumn{1}{c}{Type}& \multicolumn{1}{c}{Min} & \multicolumn{1}{c}{Max} &\multicolumn{1}{c}{Mean} &\multicolumn{1}{c}{Median}  \\  
\hline \\[-1.8ex] 
\textbf{Disclosure tweet length} & Count & 13 & 309 & 156.4 & 132 & Count & 17 & 321 & 157.83 & 130 \\ 
\textbf{Followers} & Count & 0 & 9,221,909 & 9878.7 & 928 & Count & 0 & 4,469,696 & 9929.3 & 1258  \\ 
\textbf{Friends} & Count & 0 & 297,303 & 2876.8 & 975 & Count & 0 & 130,005 & 2673,481 & 1,174\\ 
\textbf{\#tweets} & Count & 27 & 1,268,092 & 40,216 & 16,515 & Count & 11 & 1,551,139 & 34,298.2 & 13,256  \\ 
\textbf{Tweets with \#MeToo} & Count & 1 & 1,089 & 4.2 & 1 & Count & 1 & 2,025 & 6.23 & 1 \\
\textbf{Listed Count}& Count & 0 & 36,058 & 102.8 & 12 & Count & 0 & 34,892 & 109.2 & 16 \\
\textbf{Profile Description length}& Count & 13 & 309 & 156.4 & 132 & Count & 0 & 160 & 110.3 & 128 \\ 
\textbf{Account Age (years)} & Years & 18 & 3,971 & 1439.1 & 94 & Years &8 & 3,474 & 1,522 & 703  \\
\textbf{Total favorites}& Count & 0 & 1,008,226 & 32,107.4 & 10,277.5 & Count & 0 & 949,390 & 35,274 & 12,805 \\
\textbf{Disclosure length}& Count & 13 & 309 & 156.4 & 132 & Count & 0 & 321 & 157.83 & 217 \\
\textbf{Favorites}& Count & 0 & 20,224 & 6.07 & 0 & Count & 0 & 3,429 & 9.15 & 0 \\
\textbf{Retweets}& Count & 0 & 90,890 & 2.41 & 0 & Count & 0 & 321,435 & 7.80 & 0   \\

 \textbf{Verified}& Binary & NA & 216 (5\%) & NA & NA & Binary & NA & 450(7.71\%) & NA & NA\\
  \textbf{Identifiable Profile Picture}& Binary & NA & 3,198(74.02\%) & NA & NA & Binary  & NA & 4,472(81.29\%)& NA & NA\\
\hline  \hline 
\end{tabular}
}
\end{table*}

\subsection{User tweets Dataset: Collecting user timelines}
\label{collecting-timelines}

For the purposes of determining if users share more toxic tweets against their opposite gender after the start of the \#MeToo movement, we collected the user timelines for our experiments using the Twitter API~\citep{twitterapi:2020}. A Twitter user timeline refers to the chronological record of all tweets posted by a specific user. 
For \textbf{RQ1}, we collected the timelines for 10.3K male users and 16.9K female users who had disclosed their sexual harassment experiences on Twitter. The descriptive statistics of users who shared their sexual harassment experiences are illustrated in Table~\ref{descriptive-disclosure}. 
For \textbf{RQ2} and \textbf{RQ3}, obtaining nearly the timelines of 326K male and female users combined from our primary dataset using the API would have been extremely time-consuming. 
Thus, we randomly sampled 10K male and 10K female users and obtained their timelines for our analysis for \textbf{RQ2}.
We made sure there was no overlap in users between \textbf{RQ1} and \textbf{RQ2}. The descriptive statistics of these randomly sampled users  are illustrated in Table~\ref{descriptive-1}.

\begin{table}[t]
\caption{Descriptive statistics for randomly chosen Male and Female users from our dataset. }
\label{descriptive-1}
\centering 
\resizebox{\textwidth}{!}{%
\begin{tabular}{lccccc|ccccc}
\hline \\[-1.8ex] 
& \multicolumn{5}{c}{Male users (n=10,000)} & \multicolumn{5}{c}{Female users (n=10,000)} \\
\hline \hline \\[-1.8ex] 
Statistic & Type & Min & Max& Mean&Median & Type & Min & Max &Mean &Median  \\  
\hline \\[-1.8ex] 
\textbf{Followers} & Count &0 & 9,221,909 & 8920,481 &774 & Count &0 &4,469.70 &8596.3 & 1095 \\ 
\textbf{Friends} & Count & 0 & 494,168 & 3,157,693 & 895 & Count & 0 & 515,956 & 2819.6 & 1,074\\ 
\textbf{\#tweets} & Count & 23 & 1,593,697 & 4,496,150 & 3,550 & Count & 9 & 1,551,139 & 37337.75 & 5440.8 \\ 

\textbf{Tweets with \#MeToo} & Count & 1 & 1089& 3.6 & 1 & Count & 1 & 2025 & 5.49 & 1 \\

\textbf{Listed Count}& Count & 0 & 179,214 & 109.2 & 11 & Count & 0 & 234,982 & 182.6 & 17 \\
\textbf{Profile Description length}& Count & 0 & 160 & 62.4 & 37 & Count & 0 & 160 & 79.4 & 43 \\ 
\textbf{Age (years)} & Years & 4 & 5,217 & 1043.9 & 629 & Years & 0 & 160 & 79.4 & 43 \\
\textbf{Total favorites}& Count & 0 & 1,109,520 & 30,251.58 & 7909 & Count & 0 & 1,338,154 & 34,453 & 10,645 \\
 \textbf{Verified}& Binary & NA & 315 (3.15\%) & NA & NA &Binary& NA & 564 (5.64\%) & NA & NA\\
  \textbf{Identifiable Profile Picture}& Binary & NA & 5,281 (52.81\%) & NA & NA & Binary & NA & 7,239 (72.39\%& NA & NA\\
\hline  \hline 
\end{tabular} 
}

\end{table}

\subsection{Identification of toxic posts against opposite gender}
\label{algo-toxic-against}

\begin{figure}[t]
\centering
\includegraphics[width=0.6\textwidth]{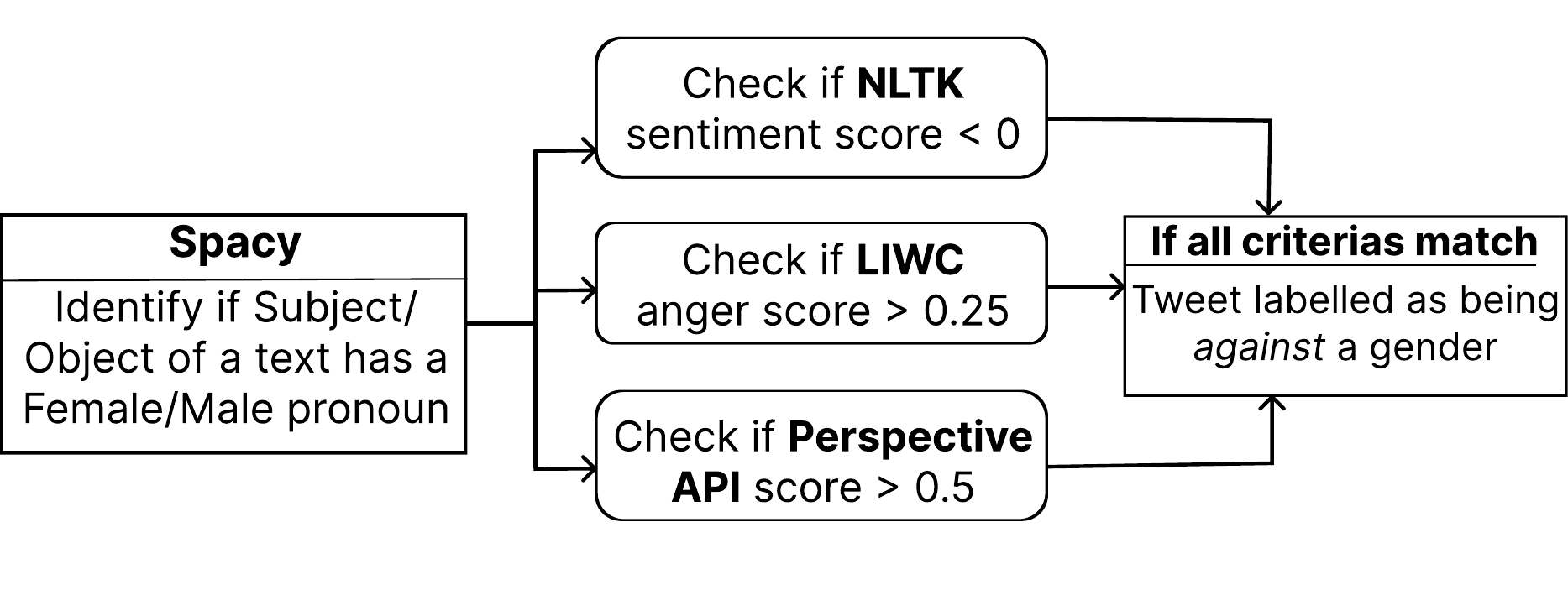}
\caption{Our framework for identifying tweets which were toxic against a particular gender}
\label{against-algo-fig}

\end{figure}

We automatically label tweets that contain toxic content against the opposite gender using an ensemble of machine learning classifiers. 
Our process uses the following tools: \textit{Firstly,} we checked if a particular tweet shared by a user was talking about women/ men by using Spacy~\citep{spacy}, an advanced open-source library for natural language processing tasks. 
Spacy uses a syntactic parser and pre-trained word embedding to recognize the subject/object tokens of a sentence, which for our purposes were female/ male pronouns and names collected from the US Census database~\citep{mislove2011understanding}. 
\textit{Secondly,} since toxic tweets do not always contain offensive words~\citep{davidson2017automated}, we use a combination of NLP-based tools to recognize the sentiment/ emotion of the tweets, which are - NLTK's sentiment analysis~\citep{nltksentiment}, Linguistic Inquiry and Word Count (LIWC)'s \cite{liwc} `Anger' score metric and Perspective API's~\citep{perspectiveapi} `Severe toxicity' score.
Our workflow for identifying gender-directed toxic tweets (also illustrated in Figure~\ref{against-algo-fig}) therefore involved: a) Identifying tweets that have men/ women pronouns as the subject/ object of the sentence and b) Having negative sentiment score($<0$) from NLTK, high anger score($>0.25$) from LIWC and high severe toxicity scores($>0.5$) from Perspective API. These thresholds were evaluated by two researchers over repeated trials over random samples, each containing 100 tweets that were flagged by the algorithm. 
Finally, the researchers manually labeled a random sample of 300 tweets, where the inter-rater agreement between the coders was 90\% with a Cohen's kappa of 0.57, indicating moderate agreement. After dissolving the disagreements, we compared the researcher labels with the labels provided by the algorithm, we found an accuracy of 87\% and a recall of 91\%, suggesting our algorithm is effective. 
The thresholds considered for each tool (NTLK, LIWC, and Perspective) were determined by the researchers to be reliable at detecting a tweet to be \textit{against} a gender. 

\subsection{Real life \#MeToo events}
In \textbf{RQ3}, we want to determine if toxic speech against men and women was impacted by pivotal real-life events of the \#MeToo movement. 
Firstly, we used a report by Chicago Tribune \cite{metoo_timeline} to identify 16 key MeToo events which occurred during the period of 15th October 2017 and 4th January 2020. We tracked toxic tweets against the opposite gender shared by the same set of users used in \textbf{RQ2} during and a week after the occurrence of each real-life \#MeToo event.
Our findings are illustrated in Section~\ref{real-life_events_analysis}.

\section{Analysis models and variables}
\label{analysis_models}
\subsection{Regression models}
In \textbf{RQ1} and \textbf{RQ2}, we examined whether the \#MeToo movement led to an increase in toxic speech against the opposite gender. Since our dependent variable i.e. the number of toxic tweets against a gender had a discrete and skewed distribution, which violates the requirements for a linear regression model, we decided to use a multivariate Poisson regression model which allowed us to analyze the relationship between the number of toxic tweets (count outcome) and the timing of the tweets (before or after the start of the \#MeToo movement) while controlling for other factors such as the characteristics of the poster's Twitter profile, their posting patterns, etc. that could influence the level of toxic speech.  We also divided our dataset into quartiles based on our dependent variable (See~\ref{dependent_variables}) and run our models on each of the quartiles. This allows us to explore potential differences across different levels of our dependent variable by checking if our predictor's effects change across these (four) quartiles.
This approach is common studies comparing gender-related outcomes~\cite{nilizadeh2016twitter} Also, examining the distribution and trends within each quartile separately allows help mitigate the influence of outliers and extreme values. 

\subsection{Dependent variables} 
~\label{dependent_variables}
The variable \emph{`No. of toxic tweets against'} denotes the number of tweets shared by a user which were identified as being toxic towards their opposite gender  (as identified by our algorithm highlighted in Section~\ref{algo-toxic-against}) during a particular timeframe, i.e., before and after the start of the \#MeToo movement. 
In \textbf{RQ2}, we used this count as a dependent variable to determine if our sample of 10K male and 10K female users tended to post more toxic tweets against their opposite gender after the start of the MeToo movement than before. 
This count is also used as a dependent variable in \textbf{RQ1} to identify whether the volume of tweets against the opposite gender that are posted by victims of sexual harassment has significantly increased after making their disclosure public. 
In \textbf{RQ1}, we also seek to find how the online community reacts to MeToo self-disclosure experiences posted by male and female users on Twitter.
\emph{Liking} or \emph{Retweeting} a tweet indicates agreement or support of it's message~\citep{boyd2010tweet,bhattacharya2014inferring}. 
Therefore, we investigate if users of one gender get significantly more likes on their disclosure tweets than others by running two separate models with no. of likes and no. of retweets received by post.

%

\subsection{Control variables} We consider several control variables in our models which can impact the dependent variables:
\newline
\textbf{Variables indicating activity:} Users participating in the \#MeToo movement with more tweets in their timeline naturally have a higher chance of  posting more tweets in opposition to their opposite gender. Thus, we considered status count as a control variable. 
Users visibility might also impact how users express themselves on Twitter, and thus we included the Followers, Friends, and Listed count~\citep{listedcount} variables as controls as well. 
Also, since users tend to acquire more visibility and persona over time, their Account age is also considered.
\newline
\textbf{Variables indicating identifiability:} Barlett et al.~\citep{barlett2016predicting} found users holding abusive or controversial opinions tend to have lesser identifiability in their accounts. 
Therefore, we considered features that can contribute to identifiability such as the length of the profile description, the presence of identifiable profile pictures, and whether the user is verified.\newline
\textbf{User involvement in the MeToo movement:} Being more involved in the MeToo movement, i.e., posting more tweets with the MeToo hashtag might indicate an opposing perception towards opposite genders, and thus we include the No. of tweets with MeToo as a control variable.

\subsection{Non-parametric Test}

In \textbf{RQ3}, our objective is to investigate the impact of real-life \#MeToo events on gender-based hostility, as manifested in the form of toxic tweets targeting the opposite gender.
Firstly, we visualize the increase of toxic tweets against men and women during pivotal events of the movement and then use the Wilcoxon rank-sum test~\cite{bu-nonparametric} to determine if the difference between the count of toxic tweets one week before, during, and one week after each event was statistically significant.
This method allowed us to quantify the changes in the volume of gender-based hostile tweets and to assess whether the \#MeToo events had a notable impact on online discussions.

\subsection{Thematic content analysis}
In \textbf{RQ4}, we investigate the content of posts containing toxic language that were influenced by the \#MeToo movement to identify the dominant themes present in these posts. 
In order to achieve this, two independent researchers conducted a topic analysis of 300 randomly selected Tweets posted during a significant real-life \#MeToo event, the arrest of American financier and convicted sex offender Jeffrey Epstein on July 6th, 2019. 
This event was chosen particularly because it was closely related to the core of the \#MeToo movement, i.e., the prevalence and impact of sexual harassment and assault, particularly in positions of power.
This focus allowed us to hone in on the dominant themes related to the core objectives of the movement, without the influence of extraneous factors.
The two researchers doing the analysis were deliberately chosen to represent both men and women, as well as different backgrounds, as the former resides in the United States and is fully aware of the details of the political situation in the country where most of the Tweets came from, and the latter is based in Europe and has an external perspective on American politics, but has an extensive understanding of the \#MeToo movement. 
The codes developed by the two researchers were discussed thoroughly and compared in order to ensure that all relevant themes emerging from the different perspectives were captured, and that all nuances were interpreted accurately. 
When using such sort of inductive methodology to build a grounded theory by identifying patterns in the data, it is inappropriate to use quantitative reliability indicators; this holds especially when the codes emerge from a process of describing the themes in the data, and the data are related to the social context and social actions with an ethical and/or political aspect, both of which are very much the case in this study ~\citep{mcdonald2019reliability}.

\subsection{Ethical considerations:} 
Throughout the course of the study, we closely monitored whether the tweets in our dataset were posted by users who maintained public accounts on Twitter.
In the event that the account access was changed to protected~\cite{twitter_protected}, we removed the user and all of their associated tweets from our dataset, and did not consider them for any further analysis.
Self disclosures of harassment and abuse incidents are traumatic for the victim, and thus to respect their privacy, we do not include the content of disclosure tweets in our work, and also do not intend to share this dataset with any third party organizations.
Only aggregated statistics of this data is used for the purposes of this paper. 
Our thematic content analysis (\textbf{Section~\ref{thematic_analysis})} directly quotes several tweets which contain offensive and undesirable language. These examples have been included to provide a comprehensive understanding of the nature and range of offensive content present on social media platforms. Reader discretion is advised. 

\section{RQ1: Sexual harassment disclosure}
\label{disclosure-analysis}
\subsection{Reception towards disclosure}
\label{disclosure_reception}

For this analysis of Research Question 1 (RQ1), we consider the self-disclosure tweets posted by the users in our dataset. The distributions for tweet favorites and retweets are highly skewed for both genders, which exhibit heavy tails. 
Thus, we employed the non-parametric Wilcoxon rank sum test to find a significant difference between the male and female groups.
Our results indicate that female users tend to get significantly more favorites on their disclosure tweets than male users do ($M_{female} =9.15$ vs. $M_{male} = 6.07$, $W = 85148884$, $p<0.001$, with high effects size). Also, female users get more retweets for their disclosure tweets than male users ($M_{female} =7.80$ vs. $M_{male} = 2.41$, $W = 85257820$, $p<0.001$, with medium effects size). 
Since the favorites and retweets variables do not follow a normal distribution, we performed Poisson regression on the whole population, as well as on the quartiles generated by dividing the dependent variables (No. of favorites, No. of retweets).

The results in Table \ref{hypothesis1} show that considering the entire sample, female users are more likely to receive a higher number of favorites on their self-disclosure tweets compared to male users ($p<0.001$). 
However, interestingly, quartile regression reveals that the effect of gender changes as one moves from the users with the least number of favorites to users with the most number of favorites. 
While in the first three quartiles, being female is an indicator of receiving more favorites, in the last quartile, the effect changes direction.
Similarly, Table \ref{hypothesis1-1} shows that considering the entire sample, female users are more likely to receive a higher number of retweets on their self-disclosure tweets compared to male users ($p<0.001$). 
However, again, quartile regression reveals that the effect of gender changes as one moves from the users with the least number of retweets to users with the most number of retweets. 
While in the first three quartiles, being female is an indicator of receiving more retweets, in the last quartile, the effect changes direction.
Our findings suggest that male users receive significantly lower engagement in the first three quartiles, in terms of both favorites and retweets, from the Twitter community towards their harassment disclosures on the platform when compared to female users. 
This might indicate that the general perception towards the male disclosures is less welcoming than those of female users, augmented by the fact that \#MeToo has often been considered a women-driven movement~\citep{martinez2021metoo}. 
However, in the fourth quartile, when comparing male and female users who have received many favorites and retweets, being a male user indicates receiving a higher number of favorites. 
This can be due to the fact that, in general, male users are more visible on Twitter ~\citep{nilizadeh2016twitter}, and having a higher level of visibility can help receive a higher number of favorites and retweets.

\begin{table}[H]
\caption{Users tend to post more tweets against their opposite gender after their self-disclosures.}
\label{hypothesis2}
\resizebox{0.8\textwidth}{!}{
\begin{tabular}{cccccc}
\hline
\multicolumn{6}{c}{\textit{Dependent variable: Total no. of tweets against}} \\ \hline \hline
After disclosure & Poisson & 0.25 Qnt. & 0.50 Qnt. & 0.75 Qnt. & 1.00 Qnt. \\ \hline
Coeffs & 1.424*** & NULL & 1.213*** & 5.606*** & 1.30*** \\ 
IRRs & 4.153*** & NULL & 3.363*** & 27.21** & 3.669*** \\ \hline
\end{tabular}
}
\end{table}

\begin{table}[t]
\caption{Female users tend to receive more favorites for their disclosure tweets in the first three quartiles.}
\label{hypothesis1}
\resizebox{0.8\textwidth}{!}{
\begin{tabular}{cccccc}
\hline
 \multicolumn{6}{c}{\textit{Dependent variable: Disclosure tweet favourites}} \\ \hline \hline
Male & Poisson & 0.25 Qnt. & 0.50 Qnt. & 0.75 Qnt. & 1.00 Qnt. \\ \hline
Coeffs &  0.383$^{***}$ & $-$0.271$^{***}$  & $-$0.385$^{***}$ & $-$0.396$^{***}$ & 0.515$^{***}$\\ 
IRRs & 1.46$^{***}$ & 0.76$^{***}$ & 0.68$^{***}$ & 0.67$^{***}$ & 1.67$^{***}$ \\ \hline
\multicolumn{6}{l}{Note: *p\textless{}0.05 ; **p\textless{}0.01 ; ***p\textless{}0.001} \\ \hline
\end{tabular}
}
\end{table}

\subsection{Victim aftermath toxic comments}
\label{disclosure_victim_toxic}

We want to determine whether users post more tweets against their opposite gender after sharing their first disclosure tweet.
We divide our data into quartiles based on the frequency of the dependent variable - "toxic tweets against," which accounts for the total number of such tweets posted by the user in their timeline, and use the Poisson regression model to identify the impact of gender on the frequency of posting toxic tweets against the opposite gender, with the temporal aspect (Before/ After posting the disclosure) also as a dependent. Table \ref{hypothesis2} presents the results. 
For brevity, we omit other control variables in our models from the table, but full tables are available upon request. To concisely compare the rate of increase/ decrease of "tweets against" based on the control variables, we utilize Incident Rate Ratios (IRR), the exponentiated coefficients of Poisson regressions. 

The model did not converge for the 1st quartile (0.25 Qnt.) due to insufficient variation in the data. 
However, we found statistical significance for the 2nd, 3rd, and 4th quartiles.
The incident rate ratio (IRR) for the 3rd quartile is 27.21\textsuperscript{**}, suggesting that users in this quartile tend to have an enormous increase of 2,621\%  in posting toxic tweets against their opposite gender after sharing their self-disclosures. 
The IRRs for the 2nd quartile and 4th quartile were 3.363\textsuperscript{***} and 3.669\textsuperscript{***} respectively, suggesting that users in this quartile tend to post 236.3\% and 266.9\% more toxic tweets against their opposite gender after sharing their self-disclosures, respectively. 

Thus overall, users tend to post significantly more toxic tweets against their opposite gender after sharing their self-disclosures. The effect is strongest in the 3rd quartile and relatively similar in the 2nd and 4th quartiles.

\begin{table}[H]
\caption{Female users tend to get more retweets for their disclosure tweets in the first three quartiles.}
\label{hypothesis1-1}
\resizebox{0.8\textwidth}{!}{
\begin{tabular}{cccccc}
\hline
 \multicolumn{6}{c}{\textit{Dependent variable: Disclosure tweet retweets}} \\ \hline \hline
Male & Poisson & 0.25 Qnt. & 0.50 Qnt. & 0.75 Qnt. & 1.00 Qnt. \\ \hline
Coeffs & $-$1.439$^{***}$ & $-$1.271$^{***}$ & $-$1.832$^{***}$ & $-$1.206$^{***}$  & 0.188$^{***}$  \\ 
IRRs & 0.237$^{***}$ & 0.28$^{***}$ & 0.16$^{***}$ & 0.29$^{***}$ & 1.21$^{***}$ \\ \hline
\multicolumn{6}{l}{Note: *p\textless{}0.05 ; **p\textless{}0.01 ; ***p\textless{}0.001} \\ \hline
\end{tabular}
}
\end{table}

\begin{table}[]
\caption{Female users with self-disclosures tend to post more tweets against women.}
\label{hypothesis2-1}
\resizebox{0.8\textwidth}{!}{
\begin{tabular}{cccccc}
\hline
 \multicolumn{6}{c}{\textit{Dependent variable: Total no. of tweets against}} \\ \hline \hline
Male & Poisson & 0.25 Qnt. & 0.50 Qnt. & 0.75 Qnt. & 1.00 Qnt. \\ \hline
Coeffs & -8.932*** & NULL & -1.389*** & -4.221*** & -2.080*** \\ 
IRRs & 0.0001*** & NULL & 2.49*** & 0.0146*** & 0.1249*** \\ \hline
\multicolumn{6}{l}{Note: *p\textless{}0.05 ; **p\textless{}0.01 ; ***p\textless{}0.001} \\ \hline
\end{tabular}
}
\end{table}

\subsection*{Female users post more toxic tweets against men after self-disclosure}

In the previous section, we established that users who disclose their harassment experiences tend to post a greater number of toxic tweets against the opposite gender. 
In this section, we investigate whether there is a difference between genders in terms of the frequency of toxic tweets posted following self-disclosures related to the \#MeToo movement. 
Our aim is to identify any potential disparities in how men and women respond to or engage with the movement after sharing their personal experiences. 
We run the same multivariate regression model on the quartiles, this time with both "toxic tweets against" and "Gender" as the dependent variables. 

Our findings are illustrated in Table \ref{hypothesis2-1}. 
Similar to the previous experiment, our model does not converge for the first quartile, but for the other quartiles, we consistently find statistical significance that female users were more likely to post tweets against men after their self-disclosure with 2nd (0.50 Qnt.), 3rd (0.75 Qnt.) and 4th (1.0 Qnt). having co-efficient values of -1.389\textsuperscript{***}, -4.221\textsuperscript{***} and -2.080\textsuperscript{***}, respectively. The incident rate ratio (IRR) for the 2nd quartile is 2.49\textsuperscript{***}, suggesting that female users in this quartile tend to post 149\% more toxic tweets against men after sharing their self-disclosures compared to male users who share toxic tweets against women. 
The IRR in the 3rd quartile was 0.0146\textsuperscript{***}, indicating female users post 98.54\% more toxic tweets in toxic tweets posted by female users compared to men. 
This statistic for the 4th quartile, based on an IRR of 0.1249\textsuperscript{***}, was 87.51\% more toxic tweets against men tweets bt women.

Thus overall, female users who have self-disclosed post significantly more toxic tweets against men compared to male users who share toxic tweets against women.
The effect is strongest in the 3rd quartile and relatively similar in the 2nd and 4th quartiles.

\section{RQ2: toxic speech against gender}
\label{toxic-opposite-gender}

\subsection{Overall toxic speech against opposite gender}
\label{toxic-overall}


\begin{table*}[H]

\caption{Descriptive statistics for randomly chosen Male and Female users from our dataset. }
\label{descriptive-1}
\centering 
\resizebox{0.85\textwidth}{!}{%
\begin{tabular}{lccccc|ccccc}
\hline \\[-1.8ex] 
& \multicolumn{5}{c}{Male users (n=10,000)} & \multicolumn{5}{c}{Female users (n=10,000)} \\
\hline \hline \\[-1.8ex] 
Statistic & Type & Min & Max& Mean&Median & Type & Min & Max &Mean &Median  \\  
\hline \\[-1.8ex] 
\textbf{Followers} & Count &0 & 9,221,909 & 8920,481 &774 & Count &0 &4,469.70 &8596.3 & 1095 \\ 
\textbf{Friends} & Count & 0 & 494,168 & 3,157,693 & 895 & Count & 0 & 515,956 & 2819.6 & 1,074\\ 
\textbf{\#tweets} & Count & 23 & 1,593,697 & 4,496,150 & 3,550 & Count & 9 & 1,551,139 & 37337.75 & 5440.8 \\ 

\textbf{Tweets with \#MeToo} & Count & 1 & 1089& 3.6 & 1 & Count & 1 & 2025 & 5.49 & 1 \\

\textbf{Listed Count}& Count & 0 & 179,214 & 109.2 & 11 & Count & 0 & 234,982 & 182.6 & 17 \\
\textbf{Profile Description length}& Count & 0 & 160 & 62.4 & 37 & Count & 0 & 160 & 79.4 & 43 \\ 
\textbf{Age (years)} & Years & 4 & 5,217 & 1043.9 & 629 & Years & 0 & 160 & 79.4 & 43 \\
\textbf{Total favorites}& Count & 0 & 1,109,520 & 30,251.58 & 7909 & Count & 0 & 1,338,154 & 34,453 & 10,645 \\
 \textbf{Verified}& Binary & NA & 315 (3.15\%) & NA & NA &Binary& NA & 564 (5.64\%) & NA & NA\\
  \textbf{Identifiable Profile Picture}& Binary & NA & 5,281 (52.81\%) & NA & NA & Binary & NA & 7,239 (72.39\%& NA & NA\\
\hline  \hline 
\end{tabular} 
}

\end{table*}
In Research Question 2 (RQ2), we investigate if users who have posted any tweet containing the \#MeToo hashtag generally tend to post more toxic tweets against the opposite gender after the start of the movement. 
We gathered a random sample of 10,000 male and 10,000 female users from our primary dataset and assessed the tweets on their timelines. 
Employing our method for identifying toxic speech directed at the opposite gender (detailed in Section~\ref{algo-toxic-against}), we successfully identified 104,021 toxic tweets against women shared by the randomly selected male users and 70,593 toxic tweets against men shared by the randomly selected female users.
It is important to note that the analyzed tweets were not required to include the \#MeToo keyword, as users may not consistently use the hashtag when posting toxic tweets directed at the opposite gender.

We divided the population into quartiles based on the number of toxic tweets shared by each user and ran a multivariate Poisson regression model of them. 
The results are presented in \textbf{Table \ref{hypothesis3}.} Our model does not converge in the 1st quartile due to a lack of variation in the data. 
However, we find statistical significance in the 2nd, 3rd, and 4th quartiles.

In the 2nd quartile (0.5 Qnt), the incident rate ratio (IRR) is 4.014\textsuperscript{***}, indicating that users in this quartile post 301.4\% more toxic tweets against the opposite gender after the \#MeToo movement. Similarly, the 3rd quartile (0.75 Qnt.) has an IRR of 7.72\textsuperscript{**}, showing a 672\% increase in toxic tweets against the opposite gender after the movement.
Finally, the 4th quartile has an IRR of 3.698, suggesting that users in this quartile post 269.8\% more toxic tweets against the opposite gender following the \#MeToo movement.

Thus overall, users generally tend to post more toxic tweets against their opposite gender following the \#MeToo movement. 
The effect is most prominent in the 3rd quartile, with a 672\% increase in toxic tweets, while the 2nd and 4th quartiles exhibit relatively similar increases of 301.4\% and 269.8\%, respectively. 


\begin{table}[t]
\caption{In general, users posted more tweets against their opposite gender after the MeToo movement.}
\label{hypothesis3}
\resizebox{0.8\textwidth}{!}{
\begin{tabular}{cccccc}
\hline
\multicolumn{6}{c}{\textit{Dependent variable: Percentage of tweets against}} \\ \hline \hline
After MeToo & Poisson & 0.25 Qnt & 0.50 Qnt & 0.75 Qnt & 1.00 Qnt \\ \hline
Coeffs & 2.088*** & NULL & 1.390*** & 4.347** & 1.308*** \\ 
IRRs & 8.068*** & NULL & 4.014*** & 7.72** & 3.698 \\ \hline
 \hline
\multicolumn{6}{l}{Note: *p\textless{}0.05 ; **p\textless{}0.01 ; ***p\textless{}0.001} \\ \hline
\end{tabular}}

\end{table}

\begin{table}[t]
\caption{Male users who participated in the MeToo movement tend to post more tweets against women.}
\label{hypothesis3-2}
\resizebox{0.8\textwidth}{!}{
\begin{tabular}{cccccc}
\hline
\multicolumn{6}{c}{\textit{Dependent variable: Percentages of tweets against}} \\ \hline \hline
Male & Poisson & 0.25 Qnt. & 0.50 Qnt. & 0.75 Qnt. & 1.00 Qnt. \\ \hline
Coeffs & 8.96** & NULL & 1.39*** & 4.34** & 1.708*** \\ 
IRRs & 7785.35*** & NULL & 4.014 & 7.67** &  5.51*** \\ \hline
\multicolumn{6}{l}{Note: *p\textless{}0.05 ; **p\textless{}0.01 ; ***p\textless{}0.001} \\ \hline
\end{tabular}}
\end{table}

\subsection{Male users post more toxic tweets since the \#MeToo movement}
\label{male-more-toxic}

We further investigate the differences in the volume of toxic tweets posted against the opposite gender by male and female users who participated in the \#MeToo movement. 
The goal is to better understand the potential disparities between the genders regarding their engagement with the movement and the extent to which they express hostility towards the opposite gender. 
To do so, we divided our population into quartiles based on the dependent variable `toxic tweets against' and also considered the independent binary variable `Gender' in our model. 

Table \ref{hypothesis3-2} illustrates our findings from running the regression model. 
Again, our model does not converge for the 1st quartile (0.25 Qnt.). But the coefficients are statistically significant for the 2nd, 3rd, and 4th quartile. 
For the 2nd quartile (0.25 Qnt), the incident rate ratio (IRR) is 4.014, suggesting that male users in this quartile tend to post 301.4\% more toxic tweets against women after the start of the \#MeToo movement, compared to women who post toxic tweets against men. 
In the 3rd quartile (0.75 Qnt.), the IRR is 7.67\textsuperscript{**}, indicating a 667\% increase in toxic tweets against women by male users after the movement's start compared to their female counterparts. 
Meanwhile, in the 4th quartile (1.00 Qnt.), the IRR is 5.51\textsuperscript{***}, showing a 451\% increase in toxic tweets against women by male users after participating in the \#MeToo movement, compared to women targeting men.

Thus overall, male users who participated in the \#MeToo movement tend to post more toxic tweets against women than women posting toxic tweets against men.
This effect is seen across the 2nd, 3rd, and 4th quartiles, with the strongest effect observed in the 3rd quartile. 

\subsection{Frequency of offensive terms in negative gender-based text}
\label{offensive-terms}
Having already established that users tend to post more toxic tweets against the opposite gender after disclosing their harassment experiences and that male users who participate in the movement tend to post more toxic tweets against women compared to female users posting toxic tweets against men, we now aim to investigate whether the \#MeToo movement has also influenced the use of derogatory language in these tweets. 
While it is important to note that derogatory language does not encompass the full scope of toxic speech, 
expanding on this additional layer of analysis allows us to delve deeper into the nuances of gender-based hostility on social media by examining the specific language choices and the frequency of offensive terms used by users.

\begin{figure}[H]
    \centering
    \begin{subfigure}{0.6\columnwidth}
        \includegraphics[width=\linewidth]{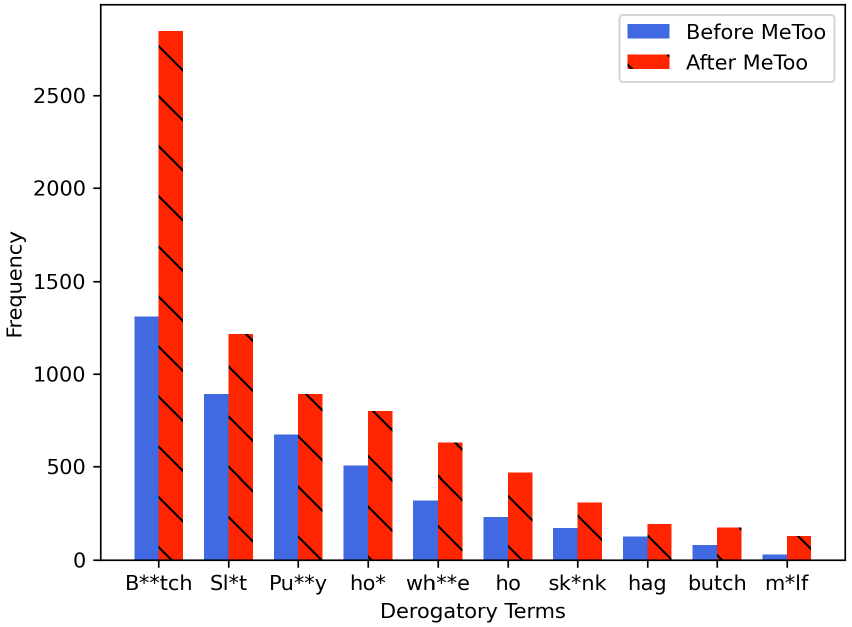}
        \caption{Distribution of the top ten offensive terms used in tweets against women by male users before and after the start of the \#MeToo movement.}
        \label{fig:offensivea}
    \end{subfigure}
    \hfill
    \begin{subfigure}{0.6\columnwidth}
        \includegraphics[width=\linewidth]{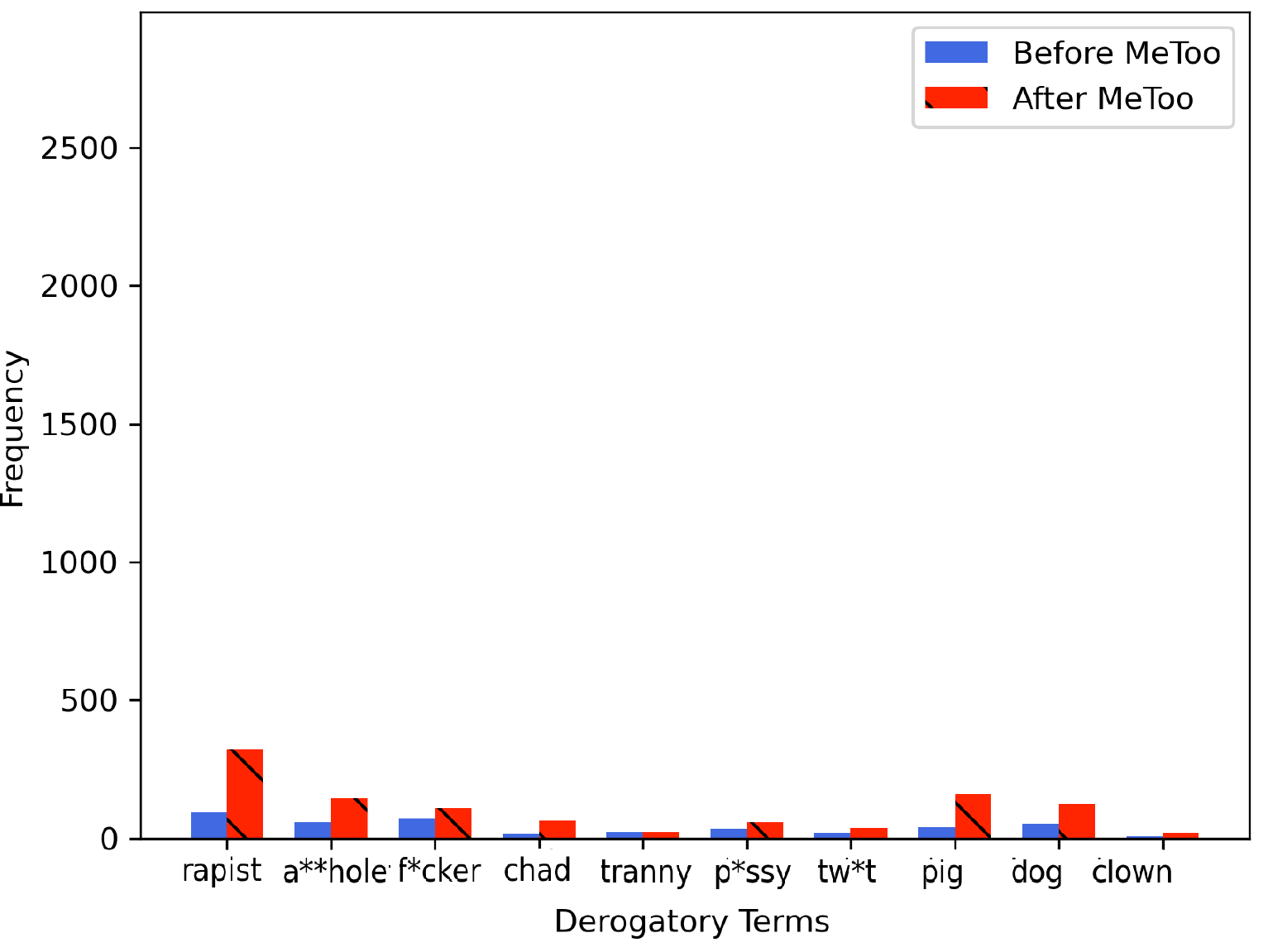}
        \caption{Distribution of the top ten offensive terms used in tweets against men by female users before and after the start of the \#MeToo movement.}
        \label{fig:offensiveb}
    \end{subfigure}
    \caption{Distributions of the top ten offensive terms used in tweets against different genders by users before and after the start of the \#MeToo movement.}
    \label{fig:offensive}
\end{figure}

We utilized 104 unique words from Hatebase~\citep{hatebase} to identify derogatory tweets targeting women. We discovered 5,701 tweets containing at least one offensive term before the \#MeToo movement and 8,392 such tweets after its inception. 
The Wilcoxon rank sum test revealed that male users were significantly more likely to employ offensive terms in their tweets after the movement began ($p<0.01$, with strong effect size) compared to before its start.
Derogatory term usage by these users increased by 67\% following the onset of the \#MeToo movement, indicating that men not only tweeted more negatively about women after October 2017 but also employed more offensive language in their tweets. 
The histogram of the ten most frequent offensive or derogatory words used by male users in their toxic tweets against women is illustrated in Figure~\ref{fig:offensivea}.

We also investigated whether female users posted more offensive tweets directed at men. For this analysis, we selected 37 unique words from Hatebase~\cite{hatebase} that were considered offensive or derogatory towards men. 
We found 487 tweets containing offensive terms targeting men before the movement and 1,308 after. 
The distribution of offensive tweets was highly skewed towards certain users, so we employed the non-parametric Wilcoxon rank sum test. 
The results showed that female users were significantly more likely to post tweets with offensive terms against men after the movement's start than before ($p<0.01$, with medium effect size). 
Overall, there was a 248\% increase in the use of offensive terms by female users following the \#MeToo movement. 
The histogram of the ten most frequent offensive or derogatory words used by female users in their toxic tweets against women is illustrated in Figure~\ref{fig:offensiveb}.


In summary, our analysis of offensive language usage in tweets highlights a substantial escalation in the use of derogatory terms by both male and female users following the start of the \#MeToo movement. 
This observation signifies that the movement has not only influenced the volume of gender-based hostility on social media platforms like Twitter, but it has also intensified the nature of such interactions, as users increasingly employ harsher language and offensive terms in their tweets against the opposite gender.

\section{RQ3: Impact of real-life events towards gender-based Toxicity}
\label{real-life_events_analysis}

\begin{table*}[H]
\centering
\caption{Major events concerning the \#MeToo movement.}
\label{me-too-real-life-events}
 \scalebox{0.8}{
\resizebox{\textwidth}{!}{%
\begin{tabular}{|c|c|c|c|c|c|}
\hline
\textbf{Number} & \textbf{Date} & \textbf{Event} & \textbf{Number} & \textbf{Date} & \textbf{Event} \\ \hline
\textbf{1} & October 15th, 2017 & \begin{tabular}[c]{@{}c@{}}Alyssa Milano encourages others to share\\ their sexual harassment story with the \\ "Me Too" hashtag.\end{tabular} & \textbf{9} & May 25th, 2018 & \begin{tabular}[c]{@{}c@{}}Hollywood producer Harvey Weinstein\\ turned himself to NY authorties after\\ being charged with rape.\end{tabular} \\ \hline
\textbf{2} & October 29th, 2017 & \begin{tabular}[c]{@{}c@{}}Accusation against Kevin Spacey in making\\ advances towards a 14 year old in the past.\end{tabular} & \textbf{10} & August 20th, 2018 & \begin{tabular}[c]{@{}c@{}}Actress Asia Argento, a prominent activist \\ of the movement, settled a complaint for \\ sexual harassment filed by a young actor.\end{tabular} \\ \hline
\textbf{3} & November 29th, 2017 & \begin{tabular}[c]{@{}c@{}}The "Today" show fires co-host Matt Lauer over\\ allegations over sexual misconduct.\end{tabular} & \textbf{11} & Sept 25th, 2018 & \begin{tabular}[c]{@{}c@{}}Bill Cosby sentenced to three to ten years\\ behind bars.\end{tabular} \\ \hline
\textbf{4} & January 1st, 2017 & \begin{tabular}[c]{@{}c@{}}More than 300 women of Hollywood form \\ "Times Up", an anti-harassment coalition.\end{tabular} & \textbf{12} & April 3rd, 2019 & \begin{tabular}[c]{@{}c@{}}Then former vice president, Joe Biden \\ promises to be "much more mindful" towards \\ respecting personal space of women.\end{tabular} \\ \hline
\textbf{5} & January 20th, 2018 & \begin{tabular}[c]{@{}c@{}}More than million people protested against\\ then President Donald Trump\end{tabular} & \textbf{13} & May 28th, 2019 & \begin{tabular}[c]{@{}c@{}}Ellen DeGeneres appears on a talk show with\\ David Letterman to talk about being sexually\\ abused by her stepfather.\end{tabular} \\ \hline
\textbf{6} & March 27th, 2018 & \begin{tabular}[c]{@{}c@{}}Former dean at MSU charged for sexual \\ harassment against female students.\end{tabular} & \textbf{14} & July 8th, 2019 & \begin{tabular}[c]{@{}c@{}}Jeffrey Epstein, a 66 year old hedge fund manager\\ faces charges for sexually abusing several underage\\  girls.\end{tabular} \\ \hline
\textbf{7} & April 6th, 2018 & \begin{tabular}[c]{@{}c@{}}Motivational speaker Tony Robbins denounces\\ the MeToo movement claiming women use it to\\ gain "victimhood".\end{tabular} & \textbf{15} & October 28th, 2019 & \begin{tabular}[c]{@{}c@{}}South California Rep. Katie Hill resigns admist\\ rumours of engaging in an inapproriate sexual\\ relationship.\end{tabular} \\ \hline
\textbf{8} & April 26th, 2018 & \begin{tabular}[c]{@{}c@{}}Bill Cosby convicted for drugging and molesting a \\ woman.\end{tabular} & \textbf{16} & January 6th, 2020 & \begin{tabular}[c]{@{}c@{}}Harvey Weinstein is indicted for new sex crime\\ charges in LA.\end{tabular} \\ \hline
\end{tabular}}
}\end{table*}

\label{impact-of-metoo_real-life-event}

\begin{figure*}[!h]
\centering
\includegraphics[width=0.8\textwidth]{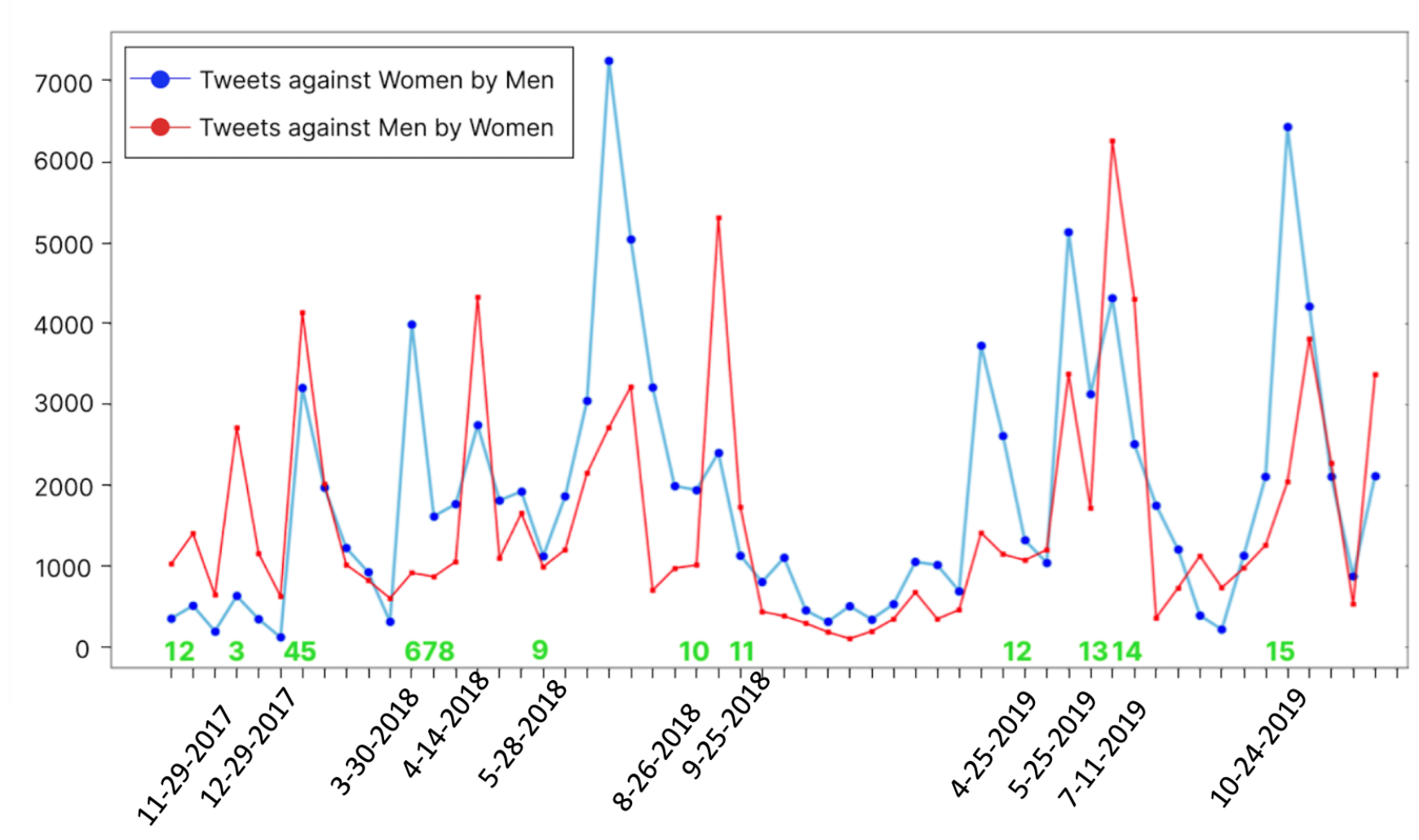}
\caption{Distribution of toxic tweets against women and against men from October 15th, 2017 to January 4th, 2020. 
Each time bin on the X-axis indicates a period of two weeks. We numerically indicate the occurrence of each real-life MeToo movement (in green) on the X-axis and for ease of interpretation, we only mark the weeks  adjacent to the occurrence of each event. 
For a brief description of events, please refer to Table \ref{me-too-real-life-events}}
\label{real-life-events}

\end{figure*}

The influence of the \#MeToo movement has been far-reaching, affecting not only online discourse but also closely intertwining with real-life events. 
To better understand the impact of these events on gender-based negativity in social media, we analyzed the connection between major \#MeToo-related events and the increase in negative tweets against men and women. 
We identified 16 key events during the period from October 15th, 2017, to January 4th, 2020, the detailed descriptions of which can be found in Table~\ref{me-too-real-life-events}. 
In Figure \ref{real-life-events}, we present the distribution of tweets against women and against men from October 15th, 2017, to January 4th, 2020. 
Each time bin on the X-axis represents a period of two weeks. 
We numerically indicate the occurrence of each real-life MeToo event (in green) on the X-axis, and for ease of interpretation, we only mark the weeks adjacent to the occurrence of each event. 

Our analysis revealed a strong correlation between these events and the frequency of negative tweets against the opposite gender. 
For instance, there was a noticeable increase in tweets against men when Kevin Spacey was accused of making advances towards a 14-year-old in October 2017~\citep{levin2017kevin}, when Matt Lauer was fired from the "Today" show over allegations of sexual misconduct in November 2017~\citep{kim2017matt}, when Bill Cosby was convicted for drugging and molesting a woman in April 2018~\citep{billcosby:2018}, and when Hollywood producer Harvey Weinstein was charged with rape in May 2018~\citep{weinsteinscandal}. 
On the other hand, there was a significant surge in tweets against women when prominent \#MeToo activist Asia Argento was accused of sexual harassment by a young actor in August 2018~\citep{asiaargento:2018}, when Ellen DeGeneres talked about being sexually abused by her stepfather in May 2019~\citep{ellendegeneres:2019} when South California Rep. Katie Hill resigned amid rumors of engaging in an inappropriate sexual relationship in October 2019~\citep{siegel2019katie}, and when Hollywood women formed the "Times Up" anti-harassment coalition in January 2018~\citep{rock2018time}.

\begin{table}[]
\centering
\caption{Major events concerning the \#MeToo movement.}
\label{me-too-real-life-events}
\resizebox{\textwidth}{!}{%
\begin{tabular}{|c|c|c|c|c|c|}
\hline
\textbf{Number} & \textbf{Date} & \textbf{Event} & \textbf{Number} & \textbf{Date} & \textbf{Event} \\ \hline
\textbf{1} & October 15th, 2017 & \begin{tabular}[c]{@{}c@{}}Alyssa Milano encourages others to share\\ their sexual harassment story with the \\ "Me Too" hashtag.\end{tabular} & \textbf{9} & May 25th, 2018 & \begin{tabular}[c]{@{}c@{}}Hollywood producer Harvey Weinstein\\ turned himself to NY authorties after\\ being charged with rape.\end{tabular} \\ \hline
\textbf{2} & October 29th, 2017 & \begin{tabular}[c]{@{}c@{}}Accusation against Kevin Spacey in making\\ advances towards a 14 year old in the past.\end{tabular} & \textbf{10} & August 20th, 2018 & \begin{tabular}[c]{@{}c@{}}Actress Asia Argento, a prominent activist \\ of the movement, settled a complaint for \\ sexual harassment filed by a young actor.\end{tabular} \\ \hline
\textbf{3} & November 29th, 2017 & \begin{tabular}[c]{@{}c@{}}The "Today" show fires co-host Matt Lauer over\\ allegations over sexual misconduct.\end{tabular} & \textbf{11} & Sept 25th, 2018 & \begin{tabular}[c]{@{}c@{}}Bill Cosby sentenced to three to ten years\\ behind bars.\end{tabular} \\ \hline
\textbf{4} & January 1st, 2017 & \begin{tabular}[c]{@{}c@{}}More than 300 women of Hollywood form \\ "Times Up", an anti-harassment coalition.\end{tabular} & \textbf{12} & April 3rd, 2019 & \begin{tabular}[c]{@{}c@{}}Then former vice president, Joe Biden \\ promises to be "much more mindful" towards \\ respecting personal space of women.\end{tabular} \\ \hline
\textbf{5} & January 20th, 2018 & \begin{tabular}[c]{@{}c@{}}More than million people protested against\\ then President Donald Trump\end{tabular} & \textbf{13} & May 28th, 2019 & \begin{tabular}[c]{@{}c@{}}Ellen DeGeneres appears on a talk show with\\ David Letterman to talk about being sexually\\ abused by her stepfather.\end{tabular} \\ \hline
\textbf{6} & March 27th, 2018 & \begin{tabular}[c]{@{}c@{}}Former dean at MSU charged for sexual \\ harassment against female students.\end{tabular} & \textbf{14} & July 8th, 2019 & \begin{tabular}[c]{@{}c@{}}Jeffrey Epstein, a 66 year old hedge fund manager\\ faces charges for sexually abusing several underage\\  girls.\end{tabular} \\ \hline
\textbf{7} & April 6th, 2018 & \begin{tabular}[c]{@{}c@{}}Motivational speaker Tony Robbins denounces\\ the MeToo movement claiming women use it to\\ gain "victimhood".\end{tabular} & \textbf{15} & October 28th, 2019 & \begin{tabular}[c]{@{}c@{}}South California Rep. Katie Hill resigns admist\\ rumours of engaging in an inapproriate sexual\\ relationship.\end{tabular} \\ \hline
\textbf{8} & April 26th, 2018 & \begin{tabular}[c]{@{}c@{}}Bill Cosby convicted for drugging and molesting a \\ woman.\end{tabular} & \textbf{16} & January 6th, 2020 & \begin{tabular}[c]{@{}c@{}}Harvey Weinstein is indicted for new sex crime\\ charges in LA.\end{tabular} \\ \hline
\end{tabular}
}
\end{table}

\subsection{Impact of Real-Life \#MeToo Events on Gender-Based Negativity Online}

Thus our findings provide valuable insights into the relationship between major real-life events associated with the \#MeToo movement and the subsequent increase in gender-based negativity on social media platforms. 
By establishing a correlation between these events and the rise in negative tweets targeting the opposite gender, we demonstrate how the online discourse both reflects and magnifies societal reactions to high-profile incidents~\citep{Dinakar2011Modeling}. 
For men, the public exposure of these incidents can intensify the perceived threat to their social status, power, and privilege~\citep{rudman2008backlash}. 
Consequently, some men may view the \#MeToo movement as an attack on their gender and engage in defensive behavior, including backlash and negative tweets against women~\citep{heskett2019metoo}. 
These reactions could be particularly pronounced during high-profile events, as men may perceive a heightened threat to their in-group identity and aim to maintain the status quo. 
Conversely, women may perceive real-life \#MeToo events as opportunities to express collective empowerment and solidarity~\citep{Trottier2014Theorising}. 
By engaging in negative tweets against men, they voice their frustration, anger, and demands for accountability regarding gender inequality and sexual harassment~\citep{mendes2019digital}. 

To further understand the complex interplay between real-life \#MeToo events and online discourse with respect to both genders, we dedicate the next section to a comprehensive qualitative analysis of a sample of these toxic posts and identify recurring themes that can provide insights into the cultural, social, and psychological aspects of gender-based negativity due to this movement. 

\section{Thematic content analysis} \label{thematic_analysis} 
\textbf{Content warning:} The following section involves analysis of potentially offensive and explicit language in tweets related to the \#MeToo movement. Please proceed with caution as it includes discussions of sexual harassment, sexism, and direct quotes from toxic tweets that may be distressing.
\newline
\newline
In this section, we qualitatively analyze a random selection of 300 toxic tweets from our dataset (150 written by male users against women, and 150 written by female users against men) shared during the week of Jeffrey Epstein's arrest (July 6th, 2019). In most cases, several themes emerged from one single Tweet, which is a sign that the views expressed were highly nuanced. The total number of themes that emerged was 39, and the themes along with their examples as well as the number of times they were identified are presented in detail in Table~\ref{metoo-qualitative-labels} and Table~\ref{metoo-qualitative-labels2}. Two of the themes appeared only in the Tweets written by women: \texttt{"Cohort with abuser"}, \texttt{"Fatigue of the public"}. The code of \texttt{"Fatigue of the public"} reflects a higher alliance of women with the movement, as they are concerned about the evolution of the movement, which has been perceived as beneficial. 
The code \texttt{"Cohort with abuser"} suggests that women would like to fully restore justice for survivors and ensure that people affiliated with abusers, often in systemic positions, who are potentially harmful to women, are also removed from power. The code \texttt{"Deceiving women"} indicates women narrating their personal experiences in the way they interpreted them. 

Eleven codes appeared only in the Tweets posted by men: \texttt{"Criticism of the movement"}, \texttt{"False accusations"}, \texttt{"Fear of being falsely accused"}, \texttt{"Resentment with the movement"}, \texttt{"Tarnishment of the movement"}, \texttt{"Injustice"}, \texttt{"Pathologizing women"}, \texttt{"Hatred towards men"}, \texttt{"Hatred towards women"}, \texttt{"Disturbing behavior"} and \texttt{"Justice for men"}. The code of \texttt{"Criticism of the movement"} emerged both from Tweets that openly doubted the majority of the accusations, e.g., \textit{"Your fake \#MeToo, lying, scheming, *ussy hat women have made every decent man a target for false, unsubstantiated accusations. Grow up or wise up"}, as well as from Tweets taking a more supportive stance towards survivors but constructively criticizing unintended ramifications of the movement or identifying some pitfalls, e.g., \textit{Nothing funny about a sexual predator like Cardi, she oughta be in prison like Bill Cosby is, but because she's a woman she gets a free pass from the \#MeToo crowd, and gets to continue making money and making young girls think that's the way to get ahead in the world"}. 
The codes \texttt{"False accusations"}, \texttt{"Fear of being falsely accused"}, \texttt{"Injustice", Pathologizing women"}, \texttt{"Hatred towards men"}, \texttt{"Revenge"}, \texttt{"Resentment with the movement"} all reflect doubt of the accusations, which depicts that these individuals deem the entire movement as invalid and see it in an entirely different light. Consequently, they appear to resent it due to feelings of threat, for example: \textit{"I agree with the male caller speaking to Shelagh Fogarty right now about the challenges raised by the \#MeToo movement. I will now ALWAYS ask for a chaperone when in a closed meeting with a single female for fears of lies or reputational damage"}. However, the codes \texttt{"Tarnishment of the movement"}, \texttt{"Disturbing behavior"} and \texttt{"Hatred Towards women"} show that some men do support the movement in general and believe the majority of the survivors, but they doubt some of the accusations and are concerned for possible defamation of the movement, e.g., \textit{"\#metoo means nothing if this continues! Too many false accusations! Innocent until proven guilty! This shows she is guilty!"}. 

The most common theme in the total dataset was \texttt{"Questioning/Criticizing political figures"} (N=78) e.g., \textit{"The audacity of Acosta invoking the \#MeToo era in defense of decisions made over a decade ago is stunning"}, which was also the most common in the subset of Tweets written by women, appearing in more than a third of the Tweets. The most common theme in the Tweets posted by men was \texttt{"Criticism of the movement"}, e.g., \textit{"Unfortunately women seem to have zero responsibility only men in the \#metoo movement..."}, which appeared in almost a third of the Tweets. Other highly common themes (N=40 or above) were \texttt{"Disappointment by the system/society"}, e.g., \textit{"This country places zero value on women and girls. They are sex objects to be taken by any man who wants them."}, which appeared mostly in Tweets posted by women, \texttt{"Negative Effects of the Movement"} e.g., \textit{"Maybe he feels that way because women like you have encouraged a lot of attention-seeking women to weaponize a certain movement called \#MeToo to destroy a man's life just by merely accusing him of something."}, which was discussed primarily by men, and \texttt{"Public Figures"} e.g., \textit{" From the entire \#metoo movement, the Aziz story is the one that should not be there."}, which was mentioned frequently both by men and women, but the rate was higher among men, and \texttt{"Hypocrisy"}, e.g. {"Womans Advocates or Political hate group.....imagine that...they care when it suits them...let that sink in"}, which was addressed mostly by men. 

The theme with the most balanced number of appearances in Tweets by men and women was \texttt{"Liar/Trying to mislead the public"}, e.g., \textit {"Does anyone really need to ask? Obviously not tRump. Everything that comes out of his mouth is lies."} (N=13 in Tweets by men and N=11 in Tweets by women). Eight of the Tweets contained disclosures of abuse/harassment, e.g., \textit{"This man touched me inappropriately and took cell phone video of me."} (N=7 by women and N=1 by men). Twenty-five of the Tweets were characterized as a mere expression of anger and rage, e.g., \textit{"I am sickened to my core....I feel like curling up and hiding. SURREAL *hit. BAD."} which was most common among women (N=20, compared to N=5 in Tweets by men).  An interesting aspect to mention is that despite choosing the Tweets from the time period around one of the major \#MeToo-related trials, many of the Tweets addressed other ongoing cases or incidents, such as that of Secretary Acosta, or the incident of a candidate representative refusing to be interviewed by a female journalist in a private space, and many others. 

The overall picture noticed through our analysis was that women are satisfied with the movement, as it brought a sense of justice for severe issues. However, they appear highly disappointed by the political figures in power, as well as the status quo, both in terms of formal institutions, such as the legal system, for example: \textit{FIRE HIM. NOW.N.J. judge who asked rape victim about closing her legs is sorry, his lawyer says} and the general attitude of society towards women: \textit{Prime example of \#MeToo being co-opted by \#WhiteFeminism to censor women's bodies. Seeing big boobs should have no bearing on whether or not someone makes the decision to assault another?}. 

Sexism was apparent across the entire dataset, in various forms, such as women and some men addressing it openly and trying to find healthier ways of interpreting situations \textit{"Stop normalizing the rape and trafficking of minors with the wrong words in your reporting"}, but also many expressing sexist views, both women \textit{"let's raise our daughters the correct way with no butt cheeks hanging out and boobs being exposed. Let's raise them with CLASS"} and men. There are also widely held concerns about the safety of children and minors, as assaults against minors were frequently addressed \textit{"Kissing anyone, even children, without consent can be considered a form of sexual assault. Sexual assault against children is considered pedophilia. Don't kiss kids without consent.}. 

Even though there were Tweets by men expressing support for women and survivors of harassment/ abuse and even criticizing the backlash against the movement \textit{"Out of control \#metoo movement? You think a woman would want to be ripped apart on national television, humiliated, embarrassed, called names by privileged old white men in Congress and the creepy perv. President, have every aspect of her life scrutinized for fun? Sure!"}, the most prominent views expressed by men were that opportunists falsely accuse many men and suffer immense consequences. Therefore men feel threatened and are cautious, which in some people's view could lead to the exclusion of women from activities \textit{"So when VP Pence says he won't have a meal with a woman without his wife present, is Pence not just behaving intelligently? Just like scientists and business leaders not wanting female interns? \#Metoo is turning into no women wanted."}. All in all, men are divided in their reactions to the movement. The men who resent the movement and express feelings of threat claim that these feelings are due to the accusations being false; with this premise, they neither openly express a view that perpetrators of harassment and abuse should not be punished nor admit that the feelings of threat come from knowing that they could potentially exhibit such behavior themselves. However, this premise is highly questionable and could be an excuse. A broader analysis that investigates the themes present in these tweets is illustrated in the next section.
\begin{landscape}
\begin{table}[!htb]

\caption{Emerging themes in toxic tweets against women and men }
\label{metoo-qualitative-labels}
\resizebox{1.4\textwidth}{!}{
\begin{tabular}{|c|c|c|c|c|ll|}
\hline
\multirow{2}{*}{\textbf{Number}} & \multirow{2}{*}{\textbf{Code}} & \multirow{2}{*}{\textbf{\begin{tabular}[c]{@{}c@{}}No of appearences in tweets\\ against women\end{tabular}}} & \multirow{2}{*}{\textbf{\begin{tabular}[c]{@{}c@{}}No of appearences in tweets\\ against men\end{tabular}}} & \multirow{2}{*}{\textbf{\begin{tabular}[c]{@{}c@{}}N of total\\ appearences\end{tabular}}} & \multicolumn{2}{c|}{\textbf{Characteristic quote}} \\ \cline{6-7} 
 &  &  &  &  & \multicolumn{1}{c|}{\textbf{Tweets against women}} & \multicolumn{1}{c|}{\textbf{Tweets against men}} \\ \hline
1 & Questioning/Criticising political figures & 26 & 52 & 78 & \multicolumn{1}{l|}{\begin{tabular}[c]{@{}l@{}}“She may have got caught up in the same blackmail scheme as the republicans. \\ Putin hacked Pelosi. She has been in Washington too long to be spotless.”\end{tabular}} & \begin{tabular}[c]{@{}l@{}}“Trump was very involved with Epstein. If given the chance prosecutors will introduce \\ damning evidence against the president. Just wait for the tweet storm to hit tomorrow \\ after charges are filed in SDNY. Trump has always been a sophomoric pervert with \\ a bully mentality.”\end{tabular} \\ \hline
2 & Disappointed with the system/society & 8 & 42 & 50 & \multicolumn{1}{l|}{\begin{tabular}[c]{@{}l@{}}“It is not a victim’s responsibility to come forward. \\ It is your responsibility to punish sex traffickers, rapists and pedophiles.”\end{tabular}} & \begin{tabular}[c]{@{}l@{}}“It's unavoidable, predatorial males enjoy the twisting of the blade! That's why they always \\ put the women on trial, to rerape her! White male system is designed to beat \\ women into their place! What do you think \#MeToo is fighting?"\end{tabular} \\ \hline
3 & Criticism of the movement & 49 & 0 & 49 & \multicolumn{1}{l|}{\begin{tabular}[c]{@{}l@{}}“Unfortunately women seem to have zero responsibility only men in the\\  \#metoo movement...I’m thinking a lil self-respect \& not over sexualization of \\ everything would go a long way...”\end{tabular}} & - \\ \hline
4 & Negative effects of the movement & 39 & 9 & 48 & \multicolumn{1}{l|}{\begin{tabular}[c]{@{}l@{}}“Watch @60mins 8.30pm and see the shocking hurt and career destruction from a false \\ \#metoo allegation of sexual assault 43 years ago. The allegator remains anonymous. \\ Her evidence blatant proven lies. John Jarratt an innocent man \& his family \\ tortured for no reason at all.”\end{tabular}} & \begin{tabular}[c]{@{}l@{}}"On the MS d-bag that won't let a woman reporter do her job. Nothing to do w. \#MeToo \\ (unless it was a windowless van). I had same issue with a male subordinate. Refused \\ coffee meeting with me in an open, crowded lobby. Later told HR he couldn't report \\ into a woman.”\end{tabular} \\ \hline
5 & Public figures & 25 & 16 & 41 & \multicolumn{1}{l|}{\begin{tabular}[c]{@{}l@{}}“From the entire \#metoo movement, the Aziz story is the one that should not be there. \\ Instead of sending """"nonverbal"""" signals, Abby could say """"no"""" and walk away.”\end{tabular}} & \begin{tabular}[c]{@{}l@{}}“Oh, still no documentary on Woody's alleged peadophilia? No metoo movement rooting \\ for his daughter who said he abused her? The grooming of his step-daughter? No calls\\  to cancel Woody \& his films?”\end{tabular} \\ \hline
6 & Hypocrisy & 32 & 8 & 40 & \multicolumn{1}{l|}{\begin{tabular}[c]{@{}l@{}}“They do a MeToo episode about sexual harassment in the workplace... ... And Gina just \\ happens to have left the show before anyone in the cast could point out, """"Hey, wasn't \\ Gina's blatant sexual objectification of Terry gross and dehumanizing?””\end{tabular}} & \begin{tabular}[c]{@{}l@{}}“Where is the so called MeToo Movement At for this pervert Jeffery Epstein. \\ But the MeToo Movement was so quick to lock up Bill Cosby. Oh that's right \\ it is because Bill Cosby was messing with a white woman that's why he is in jail. \\ What total hypocrites The so Called MeToo Is”\end{tabular} \\ \hline
7 & False accusations & 33 & 0 & 33 & \multicolumn{1}{l|}{\begin{tabular}[c]{@{}l@{}}“She wanted to ride Vic. He denied it. She went beserk being rejected. To keep gold \\ digging Ron's money, she played the role of being assaulted. To keep the show going she \\ blackmailed Funimation. """"Or you kick Vic, or I'll 'metoo' the company"""". \\ They fell for it (Spacey, Weinstein)"""\end{tabular}} & - \\ \hline
8 & Pedophilia/Abuse of minors & 3 & 29 & 32 & \multicolumn{1}{l|}{\begin{tabular}[c]{@{}l@{}}“Few of them are genuine. Take example of Asia Argento the leader of \#MeToo movement. \\ She herself admitted having sexually assaulted a minor Jimmy Bennet and paid \$3.5 million.”\end{tabular}} & \begin{tabular}[c]{@{}l@{}}"Oh for the love! UNDER AGE WOMEN?! REALLY NEWS MEDIA?? They are \\ CHILDREN, robbed of their innocence. What happened to reporting the facts? \\ Under age 18= a child. You all should be ashamed of yourselves and give these girls respect!”\end{tabular} \\ \hline
9 & Impact of sexism & 19 & 12 & 31 & \multicolumn{1}{l|}{\begin{tabular}[c]{@{}l@{}}“First the guy says he won’t allow the female to do her job because of his marriage vows, \\ then a few minutes later he says because of MeToo. This guy just doesn’t want to work with \\ women, he is a class A Misogynist."\end{tabular}} & \begin{tabular}[c]{@{}l@{}}“I agree buts let's raise our daughters the correct way with no butt cheeks hanging out and \\ boobs being exposed. Let's raise them with CLASS.”\end{tabular} \\ \hline
10 & Justice for survivors & 2 & 26 & 28 & \multicolumn{1}{l|}{\begin{tabular}[c]{@{}l@{}}“Womens soccer sucks. They got smoked by 14 year old boys. The only thing that would make\\  it better is, if they allow trans men on the team. But those sexist jerks won't do it. I LOVE \\ trans-men dominating womens sports. The \#metoo Karma monster is AWESOME!"""\end{tabular}} & \begin{tabular}[c]{@{}l@{}}"Jeffrey Epstein, KARMA sucks, doesn't it? Ocosta is out! Now it's trump's turn to\\  face his accusers. IT IS THE VICTIMS TURN AT JUSTICE!! No more covering for \\ these pedophiles! All of them.”\end{tabular} \\ \hline
11 & Exploitation of the movement & 25 & 1 & 26 & \multicolumn{1}{l|}{\begin{tabular}[c]{@{}l@{}}”She chose to make a summary judgement to build political metoo capital at the expense of \\ justice for Franken. It was pure opportunism.This is not a quality I want in a leader.He \\ should’ve gotten the investigation he requested on himself.”\end{tabular}} & \begin{tabular}[c]{@{}l@{}}"Creepy...maybe, rape? Not even in this \#MeToo era. You should apologize to all \\ who really did get raped or sexually abused. Your \#publicitystunt \#sucks""\end{tabular} \\ \hline
12 & Anger & 5 & 20 & 25 & \multicolumn{1}{l|}{“You \#MeToo wh*res are annoying”} & \begin{tabular}[c]{@{}l@{}}“It's as if he's saying the \#Metoo movement didn't exist or the public didn't care about these \\ cases therefore they gave Epstein a slap on the wrist. The law is the law regardless of \\ public perception. That's a sick excuse!!”\end{tabular} \\ \hline
13 & Liars/Trying to mislead the public & 13 & 11 & 24 & \multicolumn{1}{l|}{\begin{tabular}[c]{@{}l@{}}“The greatest trick the Devil ever pulled was convincing smart people like you that \\ Clinton was suitable to be president. This Epstein thing might just prove that. \\ Been already revealed Billy has lied about how many trips he took!”\end{tabular}} & \begin{tabular}[c]{@{}l@{}}“The @realDonaldTrump and former @SecretaryAcosta sketch is hysterical and yet \\ horribly terrifying! They‚Äôre clearly both doofy assclowns but the ease with which they\\  lie, blatantly, openly and freely is evil!”\end{tabular} \\ \hline
14 & \begin{tabular}[c]{@{}c@{}}Justification of the movement-severity \\ of issues addressed\end{tabular} & 2 & 20 & 22 & \multicolumn{1}{l|}{\begin{tabular}[c]{@{}l@{}}“You mean, in this moment of \#MeToo that rewarding someone for offering to \\ wh*re themselves out to make a terrible situation come to fruition is NOT how \\ things are supposed to work in the industry...?”\end{tabular}} & \begin{tabular}[c]{@{}l@{}}“\#MeToo is a rejection of some men’s ugly, brutal sexual power plays in the workplace \\ and on the streets. Men’s sulking resentment at having their power challenged is evident in\\  their abusive responses to women who have the audacity to stand up for \\ their basic human rights.”\end{tabular} \\ \hline
15 & Fear of being falsely accused & 21 & 0 & 21 & \multicolumn{1}{l|}{\begin{tabular}[c]{@{}l@{}}“I agree with the male caller speaking to Shelagh Fogarty right now about the \\ challenges raised by the \#MeToo movement. I will now ALWAYS ask for a \\ chaperone when in a closed meeting with a single female for fears of lies \\ or reputational damage”\end{tabular}} & - \\ \hline
16 & In cohort with abuser & 0 & 18 & 18 & \multicolumn{1}{l|}{-} & \begin{tabular}[c]{@{}l@{}}“Why Trump's White House Is Linked To Jeffrey Epstein Sex Trafficking TRUMP \\ DEAREST FRIEND FROM MANY YEARS PARTNER! IN FEISTY! ENCOUNTERS!!”\end{tabular} \\ \hline
17 & Criticism of the backlash & 9 & 9 & 18 & \multicolumn{1}{l|}{\begin{tabular}[c]{@{}l@{}}“No, in your first tweet you outright say that you think women who came \\ forward in MeToo are liars. You don’t respect women whatsoever. You think \\ traumatized assault victims who bravely speak up are making it up.”\end{tabular}} & \begin{tabular}[c]{@{}l@{}}“Foster believes women are the cause of the 'MeToo' movement. He thinks that \\ women manipulate the truth and are lying, and by forcing females to be accompanied \\ by another man, he will protect himself. NEWSFLASH: women aren't the cause of the\\  'MeToo' movement, men are."\end{tabular} \\ \hline
18 & Reformulating public perceptions & 1 & 16 & 17 & \multicolumn{1}{l|}{\begin{tabular}[c]{@{}l@{}}“disgusting. calling him the VICTIM of \#MeToo he sexually assaulted \\ someone and yet you call it a creepy date and HE’s the victim. what about the woman\\  he assaulted? f*ck all of this”\end{tabular}} & \begin{tabular}[c]{@{}l@{}}“This is sexual assault why is this not reported to police and this person arrested? \\ Forcing someone to touch your genitals and masturbating in front of them? \\ How sick are people today? Disgusting behavior.”\end{tabular} \\ \hline
19 & Value of the movement & 1 & 16 & 17 & \multicolumn{1}{l|}{\begin{tabular}[c]{@{}l@{}}“If @SenKamalaHarris is so concerned with the black community, why is her district\\  a mess? What unfulfilled promises has she made there? Any successes beyond being \\ pretty?Honesty/Integrity/Selfless: NOPE. In the age of \#MeToo, how can \\ Dems back someone who slept her way to power?"\end{tabular}} & \begin{tabular}[c]{@{}l@{}}“This decade is about accountability hey! This Jeffrey Epstein guy’s crimes have been \\ all over the net for years. I bet he felt invincible because nothing ever happened to him\\ ... UNTIL NOW. Say what you will abt the \#MeToo movement but it was the \\ catalyst for all this”\end{tabular} \\ \hline
20 & Urge to action & 4 & 13 & 17 & \multicolumn{1}{l|}{\begin{tabular}[c]{@{}l@{}}“She allegedly faked her suicide to falsely implicate her husband and in-laws. she\\  blaming husband for extreme step. She was found hale \&" " hearty in Bengaluru.\\ Women should also be arrested if they make false cases”\end{tabular}} & \begin{tabular}[c]{@{}l@{}}“Ray Diaz..THIS SICK F*CK HITS THIS 17 YEAR OLD GIRL!?! and even dated more \\ younger people LAPD PLEASE ARREST THIS DUDE HE MIGHT MAKE A MOVE \\ IF YA'LL DON'T DO ANYTHING! THROW THIS MAN IN JAIL”\end{tabular} \\ \hline
21 & Resentment towards the movement & 15 & 0 & 15 & \multicolumn{1}{l|}{\begin{tabular}[c]{@{}l@{}}“The \#metoo movement has men scared. False accusations are a usual weapons in\\  a women’s arsenal if she is that type of person. Taking away any chance of it is best”\end{tabular}} & - \\ \hline
\end{tabular}}
\end{table}
\end{landscape}

\begin{landscape}
\begin{table}[!htb]
\caption{Emerging themes in toxic tweets against women and men }
\label{metoo-qualitative-labels2}
\resizebox{1.4\textwidth}{!}{
\begin{tabular}{|c|c|c|c|c|ll|}
\hline
\multirow{2}{*}{\textbf{Number}} & \multirow{2}{*}{\textbf{Code}} & \multirow{2}{*}{\textbf{\begin{tabular}[c]{@{}c@{}}No of appearences in \\ tweets against women\end{tabular}}} & \multirow{2}{*}{\textbf{\begin{tabular}[c]{@{}c@{}}No of appearences in \\ tweets against men\end{tabular}}} & \multirow{2}{*}{\textbf{\begin{tabular}[c]{@{}c@{}}N of total\\ appearences\end{tabular}}} & \multicolumn{2}{c|}{\textbf{Characteristic quote}} \\ \cline{6-7} 
 &  &  &  &  & \multicolumn{1}{c|}{\textbf{Tweets against women}} & \multicolumn{1}{c|}{\textbf{Tweets against men}} \\ \hline
22 & Inaction & 3 & 9 & 12 & \multicolumn{1}{l|}{\begin{tabular}[c]{@{}l@{}}“You said nothing while knowing girls were being abused. You have screamed for\\  days about @realDonaldTrump and the evil you think he is doing and yet here you are telling \\ Twitter about Epstein's pimp. Hollywood knows all.And kept quiet”\end{tabular}} & \begin{tabular}[c]{@{}l@{}}“It's as if he's saying the \#Metoo movement didn't exist or the public didn't care \\ about these cases therefore they gave Epstein a slap on the wrist. The law is the \\ law regardless of public perception. That's a sick excuse!!”\end{tabular} \\ \hline
23 & Racial Inequalities & 5 & 5 & 10 & \multicolumn{1}{l|}{“Many white girls get away with false rape accusations...”} & \begin{tabular}[c]{@{}l@{}}"I can’t stand when men (mostly black) use Harvey Weinstein as a scapegoat. \\ Y’all must have terrible memory bc the \#metoo movement started because of HARVEY\\  WEINSTEIN. Sadly, we only paid attention to the assault of black girls after white\\  celebs came forward about a WHITE MAN."\end{tabular} \\ \hline
24 & Disclosure & 1 & 7 & 8 & \multicolumn{1}{l|}{\begin{tabular}[c]{@{}l@{}}"It was dark and I was on the way home after a late night. She approached\\  and started to insult me. I kept walking and started to feel uncomfortable.\\  After a while, she started to push me and few minutes later I was getting raped.”\end{tabular}} & \begin{tabular}[c]{@{}l@{}}“United allowed a drunk passenger to board. This man touched me inappropriately and \\ took cell phone video of me. United has no response other than a minimal flight voucher.\\  Boycott United!”\end{tabular} \\ \hline
25 & Disbelief in victims & 4 & 4 & 8 & \multicolumn{1}{l|}{\begin{tabular}[c]{@{}l@{}}“Out of control \#metoo movement? You think a woman would want to be ripped \\ apart on national television, humiliated, embarrassed, called names by privileged \\ old white men in Congress and the creepy perv President, have every \\ aspect of her life scrutinized for fun? Sure!"\end{tabular}} & \begin{tabular}[c]{@{}l@{}}“Stop ratcheting back women’s rights. Believe women when they say your boss assaulted \\ them. And quit lying.”\end{tabular} \\ \hline
26 & Support to women & 6 & 2 & 8 & \multicolumn{1}{l|}{\begin{tabular}[c]{@{}l@{}}“The stupid sexism. I'm not worried about being accused of assualt because \\ 1) I wouldn't do it 2) I, as a man, would likely have most of the benefit of the doubt \\ \&" " she would more than likely take the worst abuse from backlash. \\ Cause even w/ \#MeToo that's still the world we live in”\end{tabular}} & \begin{tabular}[c]{@{}l@{}}“NOTHING about Epstein raping you was your fault.You were an innocent child taken \\ advantage of and violated by a Predator. It‚Äôs ALSO not your fault you were too afraid\\  to speak sooner.You were traumatised. Massive respect for your courage now”\end{tabular} \\ \hline
27 & Tarnishment of the movement & 7 & 0 & 7 & \multicolumn{1}{l|}{\begin{tabular}[c]{@{}l@{}}“Women like her destroy women with legitimate stories.Trivializing and mocking\\  those who suffer real offenses in their day-to-day lives”\end{tabular}} & - \\ \hline
28 & Injustice & 6 & 0 & 6 & \multicolumn{1}{l|}{\begin{tabular}[c]{@{}l@{}}“Nothing funny about a sexual predator like Cardi, she oughta be in prison like \\ Bill Cosby is, but because she’s a woman she gets a free pass from the \#MeToo crowd,\\  and gets to continue making money and making young girls \\ think that’s the way to get ahead in the world"\end{tabular}} & - \\ \hline
29 & Sarcasm & 1 & 4 & 5 & \multicolumn{1}{l|}{\begin{tabular}[c]{@{}l@{}}“Eye contact with him?...how did anyone make eye contact with her and her gf?\\  I'm surprised nobody got """"metoo'd""""!"\end{tabular}} & \begin{tabular}[c]{@{}l@{}}“Why can’t men just get away with it - no consequences - like we used to? Ugh. \\ \#Metoo ruined it for everyone. And by everyone I mean men”\end{tabular} \\ \hline
30 & Supporting men & 4 & 1 & 5 & \multicolumn{1}{l|}{\begin{tabular}[c]{@{}l@{}}“The video proves she is lying. It’s okay to state that a woman is lying if that is \\ the truth. Despite the Metoo rules that every woman is to be believed regardless of\\  the evidence or lack thereof.”\end{tabular}} & \begin{tabular}[c]{@{}l@{}}“Only jerks who have no idea no to be respectful or have harmed women \& men are \\ scared of the \#MeToo movement. Most people are just decent human beings and \\ wouldn’t care”\end{tabular} \\ \hline
31 & Breaking the silence & 1 & 3 & 4 & \multicolumn{1}{l|}{“You think traumatized assault victims who bravely speak up are making it up."} & \begin{tabular}[c]{@{}l@{}}“If you have been exploited and/or sexually assaulted by Jeffrey Epstein and/or his \\ friends and colleagues reach out and call 1- 800-CALL FBI. You are not alone”\end{tabular} \\ \hline
32 & Disclosure indirect & 3 & 1 & 4 & \multicolumn{1}{l|}{\begin{tabular}[c]{@{}l@{}}“Where she checked her into a psych ward for two months because I ruined her life. \\ She attacked me after making sure I was mentally disabled, blind \&" " mute, she'd \\ known me for at least a year, she saw I was broken. Then she blamed me, making\\  herself a metoo victim."\end{tabular}} & \begin{tabular}[c]{@{}l@{}}“She was Thanking me and her sentence shocked me as she said these type of pigs are\\  everywhere and girls just keep silent because no one make us comfortable to raise it”\end{tabular} \\ \hline
33 & Critique on media & 1 & 3 & 4 & \multicolumn{1}{l|}{\begin{tabular}[c]{@{}l@{}}“In an era of lying \#metoo thots, the Billy Graham rule is a necessity.@washingtonpost :\\  We don't like it it because it keeps women honest \&" " won't let us fabricate dirt”\end{tabular}} & \begin{tabular}[c]{@{}l@{}}“Too many women have come out. But seems the \#MeToo movement has lost a lot of power. \\ Too many high ranking men are getting off \&" " think it is ok what they do. No publicity, no \\ action. How about it MEDIA? Bring back some justice.”\end{tabular} \\ \hline
34 & Pathologizing women & 3 & 0 & 3 & \multicolumn{1}{l|}{\begin{tabular}[c]{@{}l@{}}“The only revelation Christine Blasey Ford made crystal clear is that she's one very\\  scary person!Her schizophrenic affectation was beyond bizarre!What a pathetic poster \\ faux victim for the MeToo movement!”\end{tabular}} & - \\ \hline
35 & Hatred towards men & 3 & 0 & 3 & \multicolumn{1}{l|}{\begin{tabular}[c]{@{}l@{}}“if this was a WOMAN being told she was a FAT SL*T all the time online and she \\ snapped a d vented about it, society would comfort her! FEMINIST hate and want to\\  get rid of double standards but they are the DEFINITION of it”\end{tabular}} & - \\ \hline
36 & Fatigue of the public & 0 & 2 & 2 & \multicolumn{1}{l|}{-} & \begin{tabular}[c]{@{}l@{}}"I hope those who support the \#MeToo movement haven't gotten tired because this \\ epstein, trump, weinstein RAPE \&" " PEDOPHILE situation cannot be allowed \\ to disappear."\end{tabular} \\ \hline
37 & \begin{tabular}[c]{@{}c@{}}Process of overcoming \\ past problematic behaviors/\\ promoting good behaviors\end{tabular} & 1 & 1 & 2 & \multicolumn{1}{l|}{\begin{tabular}[c]{@{}l@{}}"I won't ask her ooo" " imma just lean forward to kiss her. I'm 99.9\% sure she won't \\ slap me. And if she refuses d kiss, respect and step back b4 e turn”\end{tabular}} & \begin{tabular}[c]{@{}l@{}}“And I’m all about the \#metoo. I’m happy his story came out and brought this convo \\ into light. But I don’t think anybody can say he clearly doesn’t feel bad. You don’t \\ know him. Maybe you don’t like what he said. But it’s probably really hard to say \\ the exact perfect thing.”\end{tabular} \\ \hline
38 & Hatred towards women & 2 & 0 & 2 & \multicolumn{1}{l|}{\begin{tabular}[c]{@{}l@{}}“I think she knows.She has two kids, my dad has the working theory that she might try \\ to take advantage of metoo for a lawsuit or something along those lines.\\ Thankfulky I'm a very unhorny drunk and I just find her obnoxious."\end{tabular}} & - \\ \hline
39 & Disturbing behavior & 1 & 0 & 1 & \multicolumn{1}{l|}{\begin{tabular}[c]{@{}l@{}}"Tlaib is anything but compassionate. She supports the maiming of young Muslim girls\\  in Mich. as there private parts are cut, clitoris removed. Tlaib on record backing the ritual. \\ She is a sadist. Where in the \#Metoo group?Fakes. Tlaib the Jew hater.”\end{tabular}} & - \\ \hline
40 & Revenge & 1 & 0 & 1 & \multicolumn{1}{l|}{\begin{tabular}[c]{@{}l@{}}“She wanted to ride Vic.He denied it. She went beserk being rejected. To keep gold digging\\  Ron's money, she played the role of being assaulted. To keep the show going she blackmailed \\ Funimation. """"Or you kick Vic, or I'll 'metoo' the company"""". \\ They fell for it (Spacey, Weinstein)"\end{tabular}} & - \\ \hline
41 & Deceiving women & 0 & 1 & 1 & \multicolumn{1}{l|}{-} & \begin{tabular}[c]{@{}l@{}}"Young women are lured into feeling *special* and, in a way powerful, when men with \\ wealth and power seek them out. They mistakenly think this sets them above other women. \\ That is until, it is turned and used against them. The women become sexual l abuse victims.”\end{tabular} \\ \hline
42 & Justice for men & 1 & 0 & 1 & \multicolumn{1}{l|}{\begin{tabular}[c]{@{}l@{}}“A great day for justice in this land. \#JohnJarrett found not guilty. And another MeToo\\  moron can jump off the \\ bandwagon and crawl back under the rock she came from.”\end{tabular}} & - \\ \hline
\end{tabular}}
\end{table}
\end{landscape}

\section{Discussion}
\label{discussion}

\subsection{Thematic Analysis of Gender-Based Hostility}
\label{themes-implications} 
\textbf{Content warning:} The following section involves analysis of potentially offensive and explicit language in tweets related to the \#MeToo movement. Please proceed with caution as it includes discussions of sexual harassment, sexism, and direct quotes from toxic tweets that may be distressing.
\newline
\newline
The thematic content analysis in \textbf{RQ4} elucidated the reasons behind the feelings of anger and rage of both men and women. For women, these negative feelings were confirmed to stem from, or at least be triggered by, a disappointment by the current state of matters with sexism being prevalent in all aspects of society and daily life, including the political scene. Among men, at least those who resent the movement, it became clear that such emotions are often driven by a sense of threat. This suits the social identity approach that Barron (2022) ~\citep{barron2022understanding} recommends for the interpretation of the negative stance of some men towards feminist movements; this would mean that these men feel that the structural privileges of their group are at stake, and therefore react negatively. In addition, it is crucial to keep in mind that these views were expressed by the people who express their anger online, often in an offensive manner, and they do not necessarily reflect the most commonly held views of society as a whole. At this point it is vital to stress once again the fact that men appear substantially divided, as many men either partially or fully support the movement, as illustrated by the concerns for the tarnishing of the movement. These results are overall in line with the existing qualitative literature. The power differences between abusers and survivors described by \citep{drewett2021breaking} from the first day of the movement, were reflected in our emerging theme of \texttt{}{"Disappointment by the system"}, which was better digested and more progressively expressed at the later stages of the movement, from our data originate. Our theme of \texttt{"Value of the movement"} strengthens the findings from the qualitative interviews Drewett conducted as part of her thesis. This author detected quite some sexism in the survey she conducted, which is also in line with our findings. Probably as a consequence of sexism, the clear association of toxic speech with these topics is a sign of the turbulence as societal changes are still developing. The topics detected by~\citep{goel2020understanding} are similar to those of the present study, even though in our sample the spread of the movement globally was not so prominent; the racial inequalities came up in our sample too, but not so boldly, probably because the majority of the participants of the movement are white \citep{mueller2021demographic}. In addition, the fact that we found a relatively small number of Tweets addressing racial inequalities, while many more expressed political criticism, is very congruent with the study of Mueller et al (2021) \citep{mueller2021demographic}, which clarified that this majority of white \#MeToo participants is more concerned about political matters. This strong political aspect that was the boldest in our sample is well in line with the study by Hansen \& Dolan (2022) \citep{hansen2022cross}, which showed that people supporting the republican party are most often against the \#MeToo movement and supporters of the democrats are in favor; this was highly occurring in our sample, often with some of the Tweets expressing even open support and criticism towards certain figures of the respective parties. Hansen \& Dolan (2022) highlight that the political ideology is more strongly associated with the stance towards the movement than gender, so this could explain the considerable percentage of Tweets supporting women written by men. The finding of themes around disbelief in survivors, claims of accusations being false and criticism of the movement that emerged from our sample of Tweets against women are very similar to the findings of Nutbeam \& Mereish (2022) \citep{nutbeam2022negative}, thus highlighting the magnitude and the persistence of these negative attitudes towards the movement. It is unfortunately apparent that this intense skepticism of men is not limited to the \#MeToo movement, but also to other feminist movements, such as \#March4justice \citep{barron2022understanding}. The results from our first research question on gender differences in the reception of disclosure, are not contradicting the findings of the studies by Bogen et al (2019, 2022) \citep{bogen2022sexual,bogen2019metoo} and Lowenstein-Barkai et al (2021) \citep{lowenstein2021me}; in fact we provide a possible explanation of their mixed findings, by identifying the importance of visibility of male accounts. The findings should be examined from a feminist lens acknowledging the historical oppression of women alongside a desire to achieve gender equity and the contextual reasons women or men might engage in aggressive behavior online. Prior literature has similarly found that female users disclosing harassment are reciprocated well by the Twitter community \citep{mendes2018metoo}, which encourages them to speak out more against their perpetrators and, in turn also promote a supportive environment for other women to share their own experiences ~\citep{bogen2019metoo}. Even if the likes and the retweets could be deemed as a minimally positive reaction, as it does not offer any tangible support \citep{austin2022hashtag}, it might have a positive impact on some survivors \citep{mendes2018metoo}. Moreover, maybe their traumatic experiences have led to a cognitive bias against men. The finding that highly visible male users also received positive reactions to their disclosures is consistent with that of work by ~\citep{nilizadeh2016twitter}, which shows for users in lower quartiles of visibility, being perceived as female is associated with more visibility; however, this flips among the most visible users where being perceived as male is strongly associated with more visibility. Another interpretation could be that men appear less credible when sharing such experiences \citep{douglas2011helpseeking,purnell2019me}, but if a man is more widely recognizable, his credibility might also be perceived as higher. Nonetheless, women might also be more inclined to have a negative attitude towards men, as is evident from their posts following their harassment disclosure, but also women who have not disclosed any such victimization. This might be due to the focus of \#MeToo on male abusers. Our finding that female users who post more tweets against men are both very visible and identifiable is suggesting that their negative posts are likely also well-received by the community. While this is not entirely surprising, given the traditional vilification of abusers ~\citep{corvo2003vilification}, it is concerning because it paints a grim picture of the impact on the overall perception of men. This is most evident from the significant increase in tweets with derogatory or offensive content against men, which significantly increased after the start of the movement, in comparison to before. In contrast to that, we also found that there was a significant increase of men indulging in negative and offensive behavior against women after the start of the movement. This indicates that male users in general have a negative perception of women,perhaps bolstered by the women-driven nature of the movement and indictment of several popular male public figures, and in the case of men with higher conservative traditional values, or fragile masculinity traits, an important drive at a deeper level could be a covert worry that the movement jeopardizes men's roles \citep{maricourt2022metoo,szekeres2020views,lisnek2022backlash}.Thus, to summarize, both men and women appear to be susceptible to cognitive biases on gender, a phenomenon seemingly triggered by the \#MeToo movement. This is justified up to some extent, given the severity of the issues addressed, and it could serve as a turning point for societal change \citep{risman2018gender,mendes2018metoo,austin2022hashtag}. Another point to be taken into account, is that anger is expressed more intensely in online posts \citep{brady2021social}, so the reality is possibly not as intense. The impact of the movement is also evident as toxic posts against both men and women surge especially in the period around pivotal events related to the movement. This does not mean that \#MeToo is the cause of this unhealthy behavior, but that these negative emotions were present due to the actual experiences of harassment, oppression, or fear of losing one's traditional roles, and the movement served merely as a situational motive for these emotions to be expressed, or as a reaction to these. The finding that toxic speech against women increased around events involving a male perpetrator highlights the magnitude of the existing misogyny and the highly conflicting cognitive biases and emotions, which point to a heightened necessity for a next step towards societal change in more constructive, healthier manners. The lower identifiability of accounts posting more toxic tweets since the start of the \#MeToo movement is in contrast to our observation in H2, but it is in line with studies in the field of cyberbullying finding that individuals making controversial or negative posts on OSNs prefer anonymity, fearing real world repercussions ~\citep{barlett2016predicting}.

\subsection{Gender Disparities in Response to Online Harassment Disclosures}

Our findings in \textbf{RQ1} have yielded essential insights into the dynamics of sexual harassment disclosures on social media platforms and the subsequent reactions from the online community. The findings should be examined from a feminist lens acknowledging the historical oppression of women alongside a desire to achieve gender equity and the contextual reasons women or men might engage in aggressive behavior online. Prior literature has similarly found that female users disclosing harassment are reciprocated well by the Twitter community \citep{mendes2018metoo}, which encourages them to speak out more against their perpetrators and, in turn also promote a supportive environment for other women to share their own experiences ~\citep{bogen2019metoo}. Even if the likes and the retweets could be deemed as a minimally positive reaction, as it does not offer any tangible support \citep{austin2022hashtag}, it might have a positive impact on some survivors \citep{mendes2018metoo}. Moreover,  their traumatic experiences may lead to a cognitive bias against men. 
On the other hand,  general perception of male disclosures is less welcoming than those by female users.

However, we find that highly visible male users tend to receive positive reactions to their disclosures, which is consistent with that of work by ~\citep{nilizadeh2016twitter}, which shows for users in lower quartiles of visibility, being perceived as female is associated with more visibility. Thus, it can be interpreted that men appear less credible, or are less visible, when sharing such experiences \citep{douglas2011helpseeking,purnell2019me}, but if a man is more widely recognizable, his credibility might also be perceived as higher. To address these disparities in engagement and perception of sexual harassment disclosures on social media, platforms should consider implementing measures to raise awareness and support for male victims, while continuing to provide a safe space for female victims
Encouraging empathy and understanding through public awareness campaigns or educational initiatives can help counter the less welcoming reception towards male disclosures.


\subsection{Increased Hostility Towards the Opposite Gender Following Harassment Disclosures}

Nevertheless, in \textbf{Section~\ref{disclosure_victim_toxic}} we found that users tended to post more tweets against the opposite gender after sharing their harassment experiences, with female users posting more toxic tweets against men after their disclosure. The increase in hostility from female users towards men after disclosing their harassment experiences could be attributed to several factors. One possibility is that these women may feel emboldened by the public sharing of their experiences, leading them to express their anger and frustration more openly. Additionally, the sharing of traumatic experiences can evoke strong emotions, including resentment and a desire for retribution, which may contribute to a more antagonistic stance towards the opposite gender. Another factor to consider is the collective impact of the MeToo movement, which has brought to light the pervasive nature of sexual harassment and abuse, predominantly perpetrated by men against women. This exposure may have amplified the feelings of anger and hostility among female users, leading them to adopt a more aggressive stance towards men in general. Furthermore, the support and solidarity that female users receive after sharing their experiences may reinforce their beliefs about the widespread prevalence of male-perpetrated harassment. As a result, they may generalize their negative experiences to the entire male population, leading to an increase in hostility. The increased discussions of people primarily with others from the same group is also how other scholars have interpreted the heightened polarization that characterizes the movement \citep{suovilla2020metoo}. It is essential to address this hostility by promoting empathy, understanding, and open dialogue between genders and people of different demographic groups and ideologies, especially on social media where this is practically more feasible. Encouraging more nuanced discussions around the experiences of all survivors, regardless of gender, can help create a more supportive and inclusive environment, reducing the likelihood of hostility and fostering healing for all those affected by sexual harassment.

\subsection{Rising Gender-Based Hostility in post \#MeToo}

The results of \textbf{RQ2} suggest that the \#MeToo movement has led to an increase in toxic speech against the opposite gender in general. A random sample of 10,000 male and 10,000 female users showed that there were more tweets expressing toxic speech against the opposite gender after the start of the movement. Moreover, negative associations were observed between controls such as profile description length, account age, and verified status with the percentage of toxic tweets posted against the opposite gender. This suggests that anonymity may play a role in the posting of toxic tweets by male users, who also tend to post a lower number of tweets with the \#MeToo hashtag.
Furthermore, our findings in \textbf{Section~\ref{male-more-toxic}} indicate that male users tend to express more negative views or toxic speech against women on Twitter compared to the amount of toxic speech female users directed towards men~\cite{PewResearch2017}. The high incidence rate in the entire population suggests that this trend is widespread and not limited to a specific subset of users.
The reasons behind this phenomenon could be multifaceted. One possible explanation is the backlash effect, where some male users may feel threatened by the \#MeToo movement's focus on gender equality and women's empowerment, leading them to express their discomfort and opposition through toxic speech. Additionally, research indicates that male users may be more likely to adopt an aggressive or confrontational stance in online interactions, which could contribute to the higher incidence of toxic speech against women~\citep{PewResearch2014}.
The detail that men of higher visibility receive more positive feedback could be related to the finding of Airey (2018) \citep{airey2018metoo} that men's narratives are perceived as more as more credible and rational than those of women in the context of \#MeToo, at least in the news. 
Another contributing factor could be the role of anonymity in online spaces. As mentioned earlier, negative associations were observed between controls like profile description length, account age, and verified status, suggesting that anonymity may enable some male users to post toxic speech without facing the consequences of their actions~\citep{Kendall2002}. Moreover, a study found that users with anonymous profiles tend to engage in more aggressive behavior online~\citep{Rainie2012}.
Overall, understanding the underlying motivations for this behavior is essential for developing effective strategies to address and mitigate the spread of toxic speech against women~\citep{Matias2019}. As more research continues to be conducted in this area, it is crucial to consider the broader societal implications of these findings and work towards creating a more inclusive and respectful online environment.

\subsection{Impact of Real-Life \#MeToo Events on Gender-Based Negativity Online}

Thus our findings provide valuable insights into the relationship between major real-life events associated with the \#MeToo movement and the subsequent increase in gender-based negativity on social media platforms. By establishing a correlation between these events and the rise in negative tweets targeting the opposite gender, we demonstrate how the online discourse both reflects and magnifies societal reactions to high-profile incidents.

For men, the public exposure of these incidents can intensify the perceived threat to their social status, power, and privilege. Consequently, some men may view the \#MeToo movement as an attack on their gender and engage in defensive behavior, including backlash and negative tweets against women. These reactions could be particularly pronounced during high-profile events, as men may perceive a heightened threat to their in-group identity and aim to maintain the status quo.

Conversely, women may perceive real-life \#MeToo events as opportunities to express collective empowerment and solidarity. By engaging in negative tweets against men, they voice their frustration, anger, and demands for accountability regarding gender inequality and sexual harassment~\citep{mendes2019digital}. 

To further understand the complex interplay between real-life \#MeToo events and online discourse with respect to both genders, we dedicate the next section to a comprehensive qualitative analysis of a sample of these toxic posts, and identify recurring themes that can provide insights into the cultural, social, and psychological aspects of gender-based negativity due to this movement.

\subsection{Recommendations}

Based on these findings we recommend that social media platforms use algorithms to promote posts that do not contain offensive expressions, but express views in a neutral, and ideally constructive fashion. It could probably also be helpful to add a content warning to the posts that contain derogatory expressions or use hashtags that are commonly used in toxic posts.However, these posts should not be banned completely, as this could create a bias and a distorted picture of reality, apart from being questionable in terms of potentially violating free speech rights. A general shift towards content oriented around healthier dialog could become part of the agendas of the new platforms that are now growing after the recent changes on Twitter. It would probably also be essential to carefully frame such changes in the policy of the platforms by pairing them with relevant campaigns, as an effort to limit the negative reactions from part of the male population that is skeptical towards feminist movements.

\subsection{Limitations}

\textbf{Only considering Male and Female users:}
A considerable limitation of the present study is that trans, queer, and gender-diverse individuals, in general, could not be included due to methodological complexity and also because the experiences of these individuals are often significantly different and merit more focused research. Therefore, future studies are highly recommended to delve into the social media discourse of gender-diverse people around the \#MeToo movement. 
\newline
\textbf{Demographic Information:}
Given that Twitter users are not required to disclose extensive demographic details information such as age, race, or location, the lack of such data points might limit the generalizability of our findings. For instance, understanding a user's age could reveal generational disparities in perspectives towards the movement or the other gender. Similarly, recognizing a user's race or ethnicity could impact their experiences and views on sexual harassment and assault.
\newline
\textbf{Language and Cultural Limitations:}
By primarily examining English-language tweets in our study, we may not have encompassed the full breadth of global perspectives on the \#MeToo movement. Not all attitudes, experiences, and viewpoints expressed in other languages on Twitter are captured, potentially resulting in the underrepresentation of non-English speaking regions or communities. The cultural nuances, expressed through language and region-specific idioms, can shape attitudes towards the movement and future research should consider incorporating multilingual and multicultural data to ensure a more comprehensive representation of the \#MeToo movement worldwide.
\section{Conclusion}

\sayak{Commented out the previous conclusion}
Our study offers investigates the impact of the \#MeToo movement on gender-based discourse on Twitter. 
While the \#MeToo movement's role in providing a platform for survivors and raising awaress on the issue of sexual harassment and assault on a global scale cannot be understated, our findings indicate that it has also unintentionally contributed to an increase in gender-based hostility, both from men and women. This hostility, expressed via increased toxic speech, signifies the emergence of deeper, ingrained societal issues and cognitive biases related to gender. The results of our analysis demonstrated that the \#MeToo movement evoked a variety of responses, with distinct trends along gender lines. Within the scope of \#MeToo, women discovered a platform for expressing their dissatisfaction and anger towards the prevalent sexism within society, emotions that were heightened by their personal experiences and observations of societal behavior. In contrast, certain men saw the movement as a direct challenge to their long-standing societal privileges. Crucially, this study underscores that the sentiments aired online are often from a vocal subset of individuals and don't reflect the broad spectrum of societal perspectives. In fact, our findings indicated that a substantial number of men were in support of the movement, either in part or in its entirety.

The rise of toxic speech that accompanied the movement signals a need for more balanced and constructive discourse. The movement has served as a reflection of existing societal attitudes towards gender, revealing areas in dire need of change and growth. As society moves forward, it is essential that future initiatives stemming from or related to \#MeToo foster dialogue that promotes understanding, reconciliation, and respect. This should be done by addressing the root causes of the observed hostility and providing spaces for healthier interactions. The aim should be to further the cause of gender equality, ensuring that the discourse around it does not perpetuate division, but rather unifies us in our common goal of achieving a more equitable society.

\bibliographystyle{ACM-Reference-Format}
\bibliography{refs}

\end{document}